\documentclass[fleqn,usenatbib]{mnras}

\usepackage[T1]{fontenc}
\usepackage{ae,aecompl}

\usepackage{graphicx}	% Including figure files
\usepackage{amsmath}	% Advanced maths commands
\usepackage{amssymb}	% Extra maths symbols
\usepackage[usenames,dvipsnames]{xcolor} % colours for comments
\newcommand{\Teff}{\mbox{$T_{\mathrm{eff}}$}}
\newcommand{\Ion}[2]{#1{\,\sc#2}}
\newcommand{\kms}{\mbox{$\mathrm{km\,s^{-1}}$}}
\newcommand{\gcc}{g\,cm$^{-3}$}
\newcommand{\Rwd}{\mbox{$R_{\mathrm{wd}}$}}
\newcommand{\Mwd}{\mbox{$M_{\mathrm{wd}}$}}
\newcommand{\Msun}{\mbox{$\mathrm{M}_{\odot}$}}
\newcommand{\Rsun}{\mbox{$\mathrm{R}_{\odot}$}}

\newcommand{\sdss}[3]{SDSS\,J#1$#2$#3}
\newcommand{\masy}{mas\,yr$^{-1}$}
\newcommand{\logX}[1]{\mbox{$\log[\mathrm{#1/He}]$}}
\newcommand{\logXY}[2]{\mbox{$\log[\mathrm{#1/#2}]$}}
\newcommand{\transition}[4]{\mbox{$^{#1}#2\,\rightarrow\,^{#3}#4$}}
\newcommand{\dd}{\mathrm{d}}
\newcommand{\Mcvz}{M_\mathrm{cvz}}
\newcommand{\Mast}{M_\mathrm{ast}}

\title[Cool DZ white dwarfs II]{Cool DZ white dwarfs II:
Compositions and evolution of old remnant planetary systems}

\author[M.A. Hollands]{
M.A. Hollands,$^{1}$\thanks{E-mail: M.Hollands@warwick.ac.uk (MH)}
B.T. G\"ansicke$^{1,2}$
and D. Koester,$^{3}$ 
\\
$^{1}$ Department of Physics, University of Warwick, Coventry CV4 7AL, UK \\
$^{2}$ Centre for Exoplanets and Habitability, University of Warwick, Coventry CV4 7AL, UK \\
$^{3}$ Institut f\"ur Theoretische Physik und Astrophysik,
University of Kiel, 24098 Kiel, Germany\\
}

\date{Accepted XXX. Received YYY; in original form ZZZ}

\pubyear{2017}

\begin{document}
\label{firstpage}
\pagerange{\pageref{firstpage}--\pageref{lastpage}}
\maketitle

\begin{abstract}

In a previous study, we analysed the spectra of 230 cool ($\Teff < 9000$\,K)
white dwarfs exhibiting strong metal contamination, measuring abundances for
Ca, Mg, Fe and in some cases Na, Cr, Ti, or Ni. Here we interpret these
abundances in terms of the accretion of debris from extrasolar planetesimals,
and infer parent body compositions ranging from crust-like (rich in Ca and Ti)
to core-like (rich in Fe and Ni). In particular, two white dwarfs,
\sdss{0823}{+}{0546} and \sdss{0741}{+}{3146}, which show $\logXY{Fe}{Ca} >
1.9$\,dex, and Fe to Ni ratios similar to the bulk Earth, have accreted by far
the most core-like exoplanetesimals discovered to date. With cooling ages in
the range 1--8\,Gyr, these white dwarfs are among the oldest stellar remnants
in the Milky Way, making it possible to probe the long-term evolution of their
ancient planetary systems. From the decrease in maximum abundances as a
function of cooling age, we find evidence that the arrival rate of material on
to the white dwarfs decreases by 3 orders of magnitude over a $\simeq 6.5$\,Gyr
span in white dwarf cooling ages, indicating that the mass-reservoirs of
post-main sequence planetary systems are depleted on a $\simeq1$\,Gyr
e-folding time-scale. Finally, we find that two white dwarfs in our sample are
members of wide binaries, and both exhibit atypically high abundances, thus
providing strong evidence that distant binary companions can dynamically
perturb white dwarf planetary systems. 

\end{abstract}


%%%%%%%%%%%%%%%%%%%%%%%%%%%%%%%%%%%%%%%%%%%%%%%%%%

%%%%%%%%%%%%%%%%% BODY OF PAPER %%%%%%%%%%%%%%%%%%

\section{Introduction}

Over the last two decades the study of extrasolar planetary systems has
revealed that worlds around other stars exhibit an unexpected level of
diversity, including system architecture, masses, and orbital parameters. Using
the method of transmission spectroscopy, it is now also possible to explore the
chemistry of exoplanet atmospheres. These have been found to contain atomic
\citep[e.g.][]{charbonneauetal02-1} and molecular species
\citep{swainetal08-1}, including multiple detections of water, e.g.
\citet{kreidberg14-2}, and in some cases clouds \citep{kreidberg14-1}. However,
at the present, the study of bulk exoplanetary properties is mostly limited to
measuring their masses and radii, and hence their bulk density. Exoplanet
structures and compositions based on the comparison of these measurements with
planet formation models, \citep[e.g.][]{lissaueretal11-1} are subject to two
sources of uncertainty. Firstly, mass and radius measurement errors are
typically large, and secondly the internal make-up of planets is degenerate
with respect to their bulk densities \citep{rogers+seager10-1}. While
mass/radius measurements do not directly yield information on internal
chemistry, they can be used for the broad classification of exoplanets.
However, during the last two decades it has become apparent that white dwarfs
can be employed as more direct probes of exoplanetary material compositions
\citep{zuckermanetal07-1}, which we outline throughout the remainder of this
introduction.

All known main-sequence exoplanet-hosts have masses well below 8\,\Msun\
(\href{https://exoplanetarchive.ipac.caltech.edu/cgi-bin/TblView/nph-tblView?app=ExoTbls&config=planets}{NASA
exoplanets archive}, accessed 22/02/2018) and are thus destined to become white
dwarfs. While any close-in ($\sim$1\,au) planetary bodies will be engulfed on
the giant branch during the post-main sequence evolution of their host star, a
large fraction of the outer planetary system will survive
\citep{burleighetal02-1, debes+sigurdsson02-1}, albeit on wider orbits due to
stellar mass-loss. This scenario has been considered for the Solar system
\citep{sackmannetal93-1, duncan+lissauer98-1}, with engulfment expected to
occur for objects out to $\simeq 1$\,au \citep{schroeder+connon08-1}. Mars, the
asteroid belt, the gas planets, and the Kuiper belt will continue in their
motion around the white dwarf Sun, but with their semi-major axes approximately
doubled. Furthermore, \citet{duncan+lissauer98-1} found that surviving
terrestrial planets and gas giants will remain stable for $>1$\,Gyr and
$>10$\,Gyr respectively. Full-lifetime simulations of multi-planet systems
indicate that exoplanets can remain on bound orbits for $>10$\,Gyr after
joining the white dwarf cooling-track \citep{verasetal16-1}.

During the long-lasting evolution  along the white dwarf cooling track, the
reduced dynamical stability can cause bodies on previously stable orbits to be
scattered \citep{debes+sigurdsson02-1}, where the least massive bodies such as
minor planets or asteroids are most susceptible. A planetesimal whose orbit is
severely perturbed can be ejected from the system, but in some cases its
periastron may instead venture sufficiently close to the white dwarf that tidal
forces will be capable of disrupting it \citep{jura03-1, verasetal14-1}.

Circularisation of the resulting dust forms a circumstellar debris disc. Such a
disc was first observed at the white dwarf G\,29-38
\citep{zuckerman+becklin87-1, grahametal90-1} via the detection of an infra-red
excess of flux. Since this discovery, discs have been observed at more than
forty other white dwarfs \citep{kilicetal05-1, becklinetal05-1,
gaensickeetal06-3, vonhippeletal07-1, juraetal07-2, farihietal08-1, farihi09-1,
brinkworthetal09-1, farihietal10-1, melisetal10-1, debesetal11-1,
kilicetal12-1, farihietal12-1, brinkworthetal12-1, bergforsetal14-1,
rochettoetal15-1, dennihyetal16-1, barberetal16-1}. Within the disc, dust
grains lose angular-momentum through a variety of mechanisms including the
Poynting-Robertson and YORP effects \citep{rafikov11-1, verasetal15-3} and are
eventually sublimated and subsequently accreted onto the white dwarf, resulting
in the appearance of metal lines in the stellar spectrum \citep{jura03-1,
debesetal12-1}\footnotemark. Because white dwarfs possess extremely high
surface gravities ($\log g\simeq 8$), these heavy elements sink below the
atmosphere on short timescales compared with the white dwarf age
\citep{schatzman47-1,paquetteetal86-2,koester09-1}, reverting the stellar
atmosphere to pure hydrogen or helium. Thus observations of white dwarfs
exhibiting metal lines in their photospheric spectra imply ongoing or recent
accretion from tidally disrupted planetesimals.

\footnotetext{Historically, metals in the atmospheres of some white dwarfs were
attributed to accretion from the interstellar-medium (ISM) \citep{wesemael79-1,
aannestadetal85-1}. However, the ISM hypothesis suffered from two major
inconsistencies. Firstly, the hydrogen-rich composition of the ISM
\citep{wilson+matteucci92-1} was at odds with the existence of metal-polluted
white dwarfs with hydrogen deficient, but helium dominated atmospheres
\citep{dupuisetal93-2, friedrichetal04-1}. Secondly, in their study of the
kinematic history of 15 metal-polluted white dwarfs, \citet{aannestadetal93-1}
found no correlation between atmospheric metals and dense regions of the ISM
\citep[see also][]{farihietal10-2}. Accretion of metals from remnants of
planetary systems \citep{jura03-1, debesetal12-1} naturally resolved these
issues.}

The most direct confirmation of the remnant planetary system framework has been
the detection of periodic transits in the light curve of WD1145+017 that vary
in depth, blocking at times up to 60 per cent of the light of the star
\citep{vanderburgetal15-1,gaensickeetal16-1,garyetal17-1}, caused by gas and
dust from the breakup of an exoplanetesimal orbiting near the Roche limit.
WD\,1145+017 also shows both an infra-red excess characteristic of a dust disc,
as well as metallic lines in its spectrum demonstrating a photosphere
contaminated with rocky debris \citep{xuetal16-1,redfieldetal17-1}. While only
one transiting system and tens of circumstellar debris discs are known at
present, metal contamination of varying degrees is observed at more than
25\,percent of white dwarfs \citep{zuckermanetal03-1, koesteretal14-1}, and
thus white dwarf spectroscopy is the most observationally accesible method used
to identify and study remnant planetary systems.

Following the first detailed study convincingly arguing that white dwarfs with
metal lines have accreted debris from their remnant planetary systems
\citep{jura03-1}, it was soon realised that spectral analysis of a white dwarf,
which yields atmospheric abundances, could be used to infer the composition of
the accreted parent bodies. The first anaylsis of this nature was performed by
\citet{zuckermanetal07-1} in their study of the metal-rich atmosphere of
GD\,362. Since then many metal-polluted white dwarfs have been analysed by a
number of authors in order to infer the compositions of the disrupted parent
bodies \citep{koester09-1, kleinetal10-1, zuckermanetal10-1, melisetal11-1,
koesteretal11-1, kleinetal11-1, gaensickeetal12-1, dufouretal12-1,
juraetal12-1, xuetal13-1, vennes+kawka13-1, farihietal13-1}. In some cases the
derived abundances are consistent with the accretion of material from
planetesimals having undergone differentiation, with lithospheric or core-like
compositions \citep{zuckermanetal11-1, wilsonetal15-1, kawka+vennes16-1}. At
the two white dwarfs GD\,61 and \sdss{124231.07}{+}{522626.6} the photospheric
abundances suggest that the accreted material is extremely oxygen rich in
comparison to the other metals, significantly exceeding the abundance expected
from the accretion of metal oxides only \citep{farihietal11-1, farihietal13-1,
raddietal15-1}. The most plausible explanation for the detected oxygen excesses
are high water fractions in the disrupted planetesimals, much like the large
Solar system asteroid Ceres. Until recently all metal-polluted white dwarfs
have shown abundance-patterns consistent with an origin from within the snow
line. However, \citet{xuetal17-1} recently demonstrated that WD\,1425+540 has
an atmosphere enriched in the volatile elements C, N, and S, and an abundance
pattern that is overall consistent with a Kuiper-belt like object. In summary,
the planetary systems around stellar remnants are found to be as diverse as
those around main-sequence stars, including our own Solar system\footnotemark.

\footnotetext{Recent review articles on the planetary systems of white dwarfs
by \citet{jura+young14-1}, \citet{farihi16-1}, and \citet{veras16-1} provide
thorough introductions to the field.}

These detailed studies of individual white dwarfs have been instrumental in
increasing our understanding of the formation and evolution of exoplanetary
systems. To date, there have been only limited attempts at the abundance
analysis of large samples of metal-rich white dwarfs, which have been typically
restricted to the measurement of Ca abundances
\citep{koesteretal05-2,dufouretal07-2,kepleretal15-1,kepleretal16-1}.

In \citet{hollandsetal17-1} (hereafter Paper~I), we presented a large sample of
231 cool white dwarfs with strong metal lines where a full abundance analysis
was performed for each object. In Paper~I, we gave particular emphasis to our
methods developed to identify these cool white dwarfs, their atmospheric
analysis, and some global properties of our sample. Here we further investigate
the detailed bulk compositions of the accreted planetesimals derived in Paper~I
and discuss the evolution of planetary systems far beyond the main-sequence. In
Section~\ref{sample}, we briefly describe our white dwarfs sample. In Section~
\ref{composition}, we present our compositional analysis of exoplanetesimals,
and discuss in detail a few unique systems in Section~\ref{extreme}.
Section~\ref{ages} makes use of the wide range of white dwarf ages to
investigate the evolution of remnant planetary systems. Finally, we present our
conclusions in Section~\ref{conclusions}.

\section{The DZ white dwarf sample}
\label{sample}

DZ white dwarfs are characterised by having \emph{only} metal lines in their
spectra \citep{sionetal83-1}. Most known DZ stars have atmospheres dominated by
helium, as helium lines become negligible below $\Teff\ \simeq 11\,000$\,K, and
as the low opacity of helium results in metal lines that are both deep and
broad even at low abundances of $\logX{Z} \lesssim -9$\,dex. In
hydrogen-dominated atmospheres with $\Teff\lesssim5000$\,K the Balmer also
lines vanish, and, if contaminated by metals, they too may be classed as DZs.
In practice these are rarely observed, as the higher opacity of hydrogen
atmospheres requires larger metal abundances to form detectable spectral lines.
Regardless of background element, DZ white dwarfs are cool objects by
definition. Because white dwarf cooling is relatively well understood
\citep{fontaineetal01-1,salaris09-1}, their effective temperatures can be used
to estimate the white dwarf ages.

In Paper~I, we identified 231 DZ white dwarfs, primarily making use of the
spectroscopy available in DR12 of the Sloan Digital Sky Survey (SDSS)
\citep{yorketal00-1,alametal15-1}. All objects in this sample exhibit strong
metallic features in their spectra with \Teff\ spanning 4500 to 9000\,K. The
relation between \Teff\ and white dwarf age is non-linear with these \Teff\
corresponding to cooling ages of $\simeq1$--$8$\,Gyr. This sample offers a
unique opportunity to investigate remnant planetary systems for a wide range of
system ages, from $1$\,Gyr up to nearly the age of the Galactic disc, providing
insight into the late-time evolution of planetary systems.

\begin{figure*}
  \centering
  \includegraphics[angle=0,width=\textwidth]{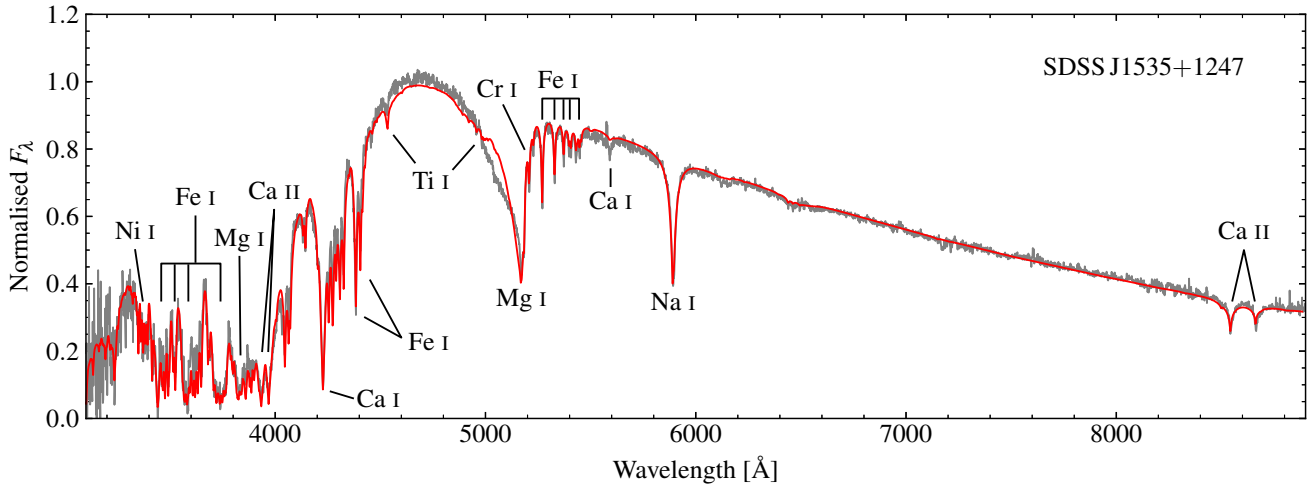}
  \caption{\label{fig:spectra_1535}
  \sdss{1535}{+}{1247} is typical of the objects in our sample both in terms of
  spectrum and composition of the accreted material, albeit  with the highest
  signal-to-noise ratio. The main spectral features are labelled. The many
  unlabelled lines in the range 4000--4400\,\AA\ are almost all from
  \Ion{Fe}{i}, with the exception of a \Ion{Cr}{i} triplet centred on
  4275\,\AA. 
  }
\end{figure*}

In Fig.~\ref{fig:spectra_1535} we show the spectrum of the DZ white dwarf
\sdss{1535}{+}{1247} with the major metal transitions labelled (see Paper~I,
Table 3 for the full precision SDSS names). As the closest member of our sample
to the Sun ($19.4\pm2.4$\,pc, Paper~I), it is sufficiently bright to allow the
detection of lines from all elements we encounter across the entire sample. The
strongest features in the DZ spectra are almost always the \Ion{Ca}{ii} H+K
resonance lines which are visible over the full range of \Teff\ in our sample.
Ca is therefore the most reliably measured element with abundances spanning
$-10$ to $-7$\,dex. Other strong Ca features include the 4227\,\AA\ \Ion{Ca}{i}
resonance line, and the infra-red \Ion{Ca}{ii} triplet (although its bluest
component is rarely strong enough to be visible). The next most easily detected
elements are Mg and Fe, whose lines have lower oscillator strengths, but are
typically $\simeq1$\,dex higher in abundance than for Ca. Fe is constrained
mostly by a forest of unresolved lines in the 3400--3900\,\AA\ region, but also
by multiplets near 4400\,\AA\ and 5400\,\AA. The \Ion{Mg}{i}-b triplet, centred
on $5171$\,\AA, is the most prominent Mg feature, especially because of its
asymmetric profile which results from a satellite line formed in high density
helium atmospheres (\citealt{allardetal16-1}, Paper~I). For 101 stars in our
sample, Na is detected via the D-doublet centred on 5893\,\AA, although the two
components are unresolved in our spectra due to pressure broadening by the
Stark effect and neutral He atoms. Cr and Ti are detected in only 60 and 27
systems respectively due to their typically low mass fractions within the
accreted material. Finally, Ni is only detected for eight objects where we have
William Herschel Telescope (WHT) follow-up spectra\footnote{A full description
of our WHT spectra and their reduction are given in Paper~I.}. These WHT
observations go as blue as 3100\,\AA, and reveal, in some cases, a set of Ni
lines centred on 3390\,\AA.

Around 10 percent of our sample have magnetic fields, detected via Zeeman
splitting, which if exceeding 1\,MG reduce the accuracy of our spectral fits
(which use non-magnetic models). These objects are tabulated in Paper~I. One of
the 231 DZ stars was sufficiently magnetic ($B \ga 20$\,MG) to prohibit a
meaningful fit to its spectrum. Thus, the following sections of this paper
discuss the remaining 230 systems.

\section{Compositions of exoplanetesimals}
\label{composition}

Relative metal abundances, obtained through spectral fitting, inform us on the
bulk compositions of the accreted planetesimals. As described in the previous
section, we are able to measure the abundances of multiple elements for 230
white dwarfs, allowing us to probe the composition of a large sample of
planetesimals formed many Gyr ago. We were able to estimate Ca, Mg, and Fe
abundances for all 230 stars in our sample, providing a common set of elements
to work with. Conveniently Ca, Mg, and Fe can be used as tracers of crust-,
mantle-, and core-like material respectively. Therefore systems that are overly
abundant in one of these elements compared to the mean, can be presumed to have
accreted material from (fragments of) differentiated planetesimals.

As mentioned in Section~\ref{sample}, Na, Cr, Ti, and Ni are not detected at
every system. However we show, particularly throughout Section~\ref{extreme},
that these elements serve as important diagnostics for confirming the nature of
material in the most extreme outliers in the Ca/Mg/Fe abundance parameter
space. 

\subsection{Relative diffusion}
\label{reldiff}

An important caveat that must be acknowledged before further discussion on
parent body compositions, is the relative diffusion of elements. In white
dwarfs with radiative atmospheres, i.e. warm DA stars, the sinking time-scales
can be as short as days to years. As this is significantly shorter than the
estimated duration of an accretion episode ($10^4$ to $10^6$\,yrs,
\citealt{girvenetal12-1}), the assumption of accretion-diffusion equilibrium in
the analysis of the accreted debris abundances is fully justified. Diffusion
time-scales vary by element with a (non-trivial) dependence on elemental
weights, meaning that the atmospheric abundance ratios do not directly
represent the composition of the accreted material \citep{paquetteetal86-2}.
However, with calculated gravitational settling time-scales in hand, this is
simple to correct for \citep{koester06-1,koester09-1}.

In contrast, in the cool, dense helium atmospheres of DZs discussed here, the
assumption of diffusion-accretion equilibrium does not hold true, as the
envelopes of these stars are unstable to convection. The outer convection zones
(CVZs) extend deep below the stellar atmosphere, and for the objects under
consideration here, contain $10^{-6}$ to $10^{-5}$ of the total white dwarf
mass. The material, once accreted into the photosphere, is diluted by
convective mixing throughout the envelope, and settles out into the core on the
diffusion time-scale at the base of the CVZ. Consequently, the depletion of
metals from the photosphere is strongly impeded. For our sample of cool white
dwarfs, these diffusion times span several $10^5$ to a few $10^6$\,yr (Table~7
of Paper~I), comparable to, or longer than the estimated duration of the
accretion phase \citep{girvenetal12-1}\footnote{For an intuitive illustration
of the very different sinking time-scales in DAs and non-DAs as a function of
white dwarf age, see Fig.~1 of \citet{wyattetal14-1}.}.

For the majority of DZs it is likely the case that accretion has ceased by the
time we observe them, with spectroscopic metals serving as an indicator of at
least one accretion event having occurred in the last few Myr. Optically thick
discs with sufficient inclination ought to be detectable around the nearest and
brightest DZs. However, so far only one DZ is known to possess an infra-red
excess indicative of a dust disc \citep{bergforsetal14-1}. This suggests that
many or most of the objects in our sample are not currently accreting, but
merely preserve the remains of former planetesimals in their photospheres for a
few Myr.

Once accretion has stopped, elements heavier than helium continue their slow
diffusion out of the CVZ, remaining spectroscopically visible for several
sinking time-scales. However the differences in diffusion velocity for each
element leads to the abundance ratios changing over time. This process is
discussed by \citet{koester09-1} in the context of GD\,362, which is well
demonstrated by their figures 2 and 3.

In the absence of accretion, the rate of change in the abundance ratio between
two elements has a simple dependence on their diffusion time-scales. For an
element $Z$ with diffusion time-scale $\tau$, its atmospheric abundance (with
respect to He) proceeds with time, $t$, as

\begin{equation}
  Z(t)/\mathrm{He} = Z(0)/\mathrm{He} \times \exp{(-t/\tau)},
  \label{eq:diff1}
\end{equation}
or in logarithmic form (base 10)
\begin{equation}
  \log{[Z/\mathrm{He}]}(t) = \log{[Z/\mathrm{He}]}(0) - \ln(10)\,t/\tau.
  \label{eq:diff2}
\end{equation}
If we consider two elements, $Z_1$ and $Z_2$, with respective diffusion
time-scales, $\tau_1$ and $\tau_2$, manipulation of either
equation\,\eqref{eq:diff1} or \eqref{eq:diff2}, leads to the relation
\begin{equation}
  \frac{\dd\log{[Z_2/Z_1]}}{\dd\log{[Z_1/\mathrm{He}]}} = \frac{\tau_1}{\tau_2} - 1.
  \label{eq:diff3}
\end{equation} 
The two important cases we consider for the remainder of this paper are Fe vs.
Ca, and Mg vs. Ca. In the first case, we find from our envelope calculations
(Paper~I) that $\tau_\mathrm{Fe}$ is usually within $5$ percent of
$\tau_\mathrm{Ca}$, and so \logXY{Fe}{Ca} effectively remains constant with
decreasing \logXY{Ca}{He}, i.e. with increasing time since the end of an
accretion episode. In the latter case, we find $\tau_\mathrm{Mg}$ is typically
a factor $2.8\pm0.1$ larger than $\tau_ \mathrm{Ca}$ for the white dwarfs in
our sample. Therefore, equation~\eqref{eq:diff3} shows that for every one\,dex
decrease in \logXY{Ca}{He}, the value of \logXY{Mg}{Ca} \emph{increases} by
$0.64$\,dex, i.e. a factor four. \citet{kawka+vennes16-1} present a similar
expression to equation~\eqref{eq:diff3}, although defined in terms of time
since the end of accretion rather than one of the absolute abundances as we do
here.

\subsection{Abundance analysis of Ca, Mg, and Fe}
\label{camgfe}

\begin{figure*}
  \centering
  \includegraphics[angle=00,width=0.8\textwidth]{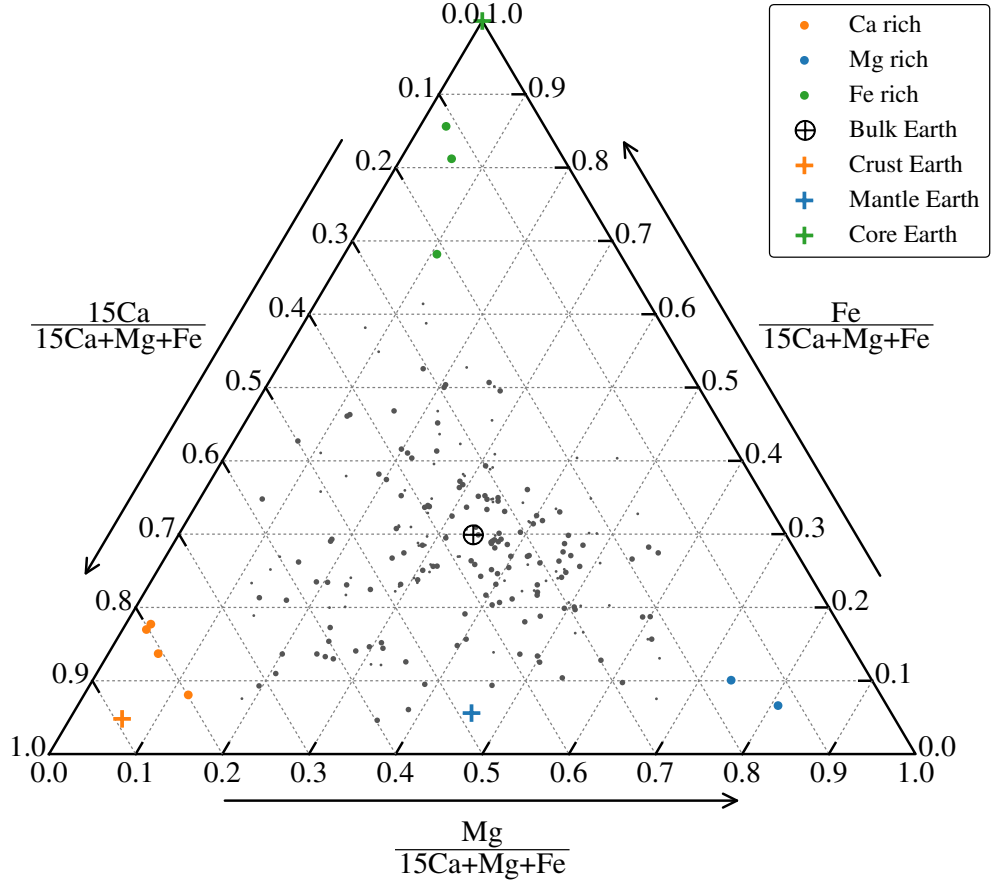}
  \caption{\label{fig:ternary}
  Ternary diagram of the Ca, Mg, and Fe abundances in our white dwarf sample,
  as measured from the spectroscopic fits. On average the Mg/Fe abundance ratio
  is found to be $\simeq 1$, whereas the Ca/Fe and Ca/Mg ratios are typically
  an order of magnitude lower. For display purposes Ca is therefore scaled by a
  factor 15, and consequently numbers on the axes do not correspond to relative
  abundance fractions. Outliers particularly rich in one of the three elements
  appear closer to the corners, where the highlighted systems are discussed in
  detail in Section~\ref{extreme}. The remaining DZs are categorised into
  higher and lower quality measurements indicated by the larger and smaller
  grey points respectively. Ticks indicate the direction along which to read
  off axis values.}
\end{figure*}

Since it is not possible to know how long ago accretion may have stopped we, at
least initially, treat the observed abundances as representative of the parent
bodies. For the three elements whose abundances we most reliably measure (Ca,
Mg, and Fe), their combined composition has two degrees of freedom, and is thus
amenable to being displayed graphically. When dealing with atomic abundances, a
typical approach is the comparison of the log-abundance ratios, e.g.
\logXY{Ca}{Mg} vs. \logXY{Fe}{Mg}. As one element (Mg in the example) appears
twice, the resulting distribution is guaranteed to contain a strong
correlation, making the positions of chemically intriguing outliers less
obvious. For this reason we display our compositions using a ternary diagram in
Fig.~\ref{fig:ternary}, where we use absolute abundances rather than the
logarithmic ratios. Because Ca is typically much less abundant than either Mg
or Fe, we rescale Ca by a factor 15 to centre the distribution within the plot
(otherwise the data appear compressed within a stripe along the right edge).
Our model atmosphere fitting method leads to some minor quantisation which is
visually distracting, we therefore initially re-sample the atmospheric
abundances with normally distributed deviates with standard deviation 0.01\,dex
(smaller than the estimated uncertainties which are at best 0.05\,dex) to
remove this artefact.

The bulk Earth composition \citep{mcdonough00-1}, $\oplus$, is located close to
the mean of our sample ($\mathrm{Ca}=0.38$, $\mathrm{Mg}=0.33$,
$\mathrm{Fe}=0.29$). This suggests that Gyr-old exoplanetesimals are overall
similar in composition to the bulk Earth as found in the analyses of younger
metal-polluted white dwarfs \citep{kleinetal10-1, kleinetal11-1,
gaensickeetal12-1, xuetal14-1}.

Typical uncertainties for the photospheric abundances measured from fitting the
white dwarf spectra are estimated to be in the range $0.05$--$0.3$\,dex. For
the poorest quality fits (low signal-to-noise spectra and low abundances), this
can translate to large scatter within Fig.~\ref{fig:ternary}, with its extent
amplified at the centre of the plot. For uncorrelated and identical Ca/Mg/Fe
uncertainties, the error in the position in dimensionless plot units is
approximately half that of the abundance errors in dex at the centre of the
diagram. For example, 0.2\,dex uncertainties on Ca/Mg/Fe abundances translates
to a positional error of 0.1 at the centre of the figure. Due to the non-linear
mapping between abundances and coordinates in the ternary diagram, the
positional errors are vastly decreased towards the corners, and therefore
systems located in these three regions represent compositionally unusual
objects even if their abundance uncertainties are large and statistically
independent. As the spectral signal-to-noise ratio is increased, the abundance
errors of Ca/Mg/Fe become positively correlated and so the uncertainties in
relative abundances, e.g. \logXY{Fe}{Ca}, become smaller still, translating to
positional uncertainties of a few $0.01$ even at the centre of the diagram.

To increase the visual weight of the systems within Fig.~\ref{fig:ternary} with
more precise spectral fits, they are displayed by the larger points, with the
smaller points corresponding to poorer measurements. We set the threshold for a
``good-quality'' fit as a median spectral signal-to-noise ratio larger than
five, and a geometric mean abundance, defined as
$(\logX{Ca}+\logX{Mg}+\logX{Fe})/3$, of at least $-8.7$\,dex. 

The most extreme systems found towards the corners of Fig.~\ref{fig:ternary}
exhibit atmospheric compositions that are Ca-rich (red), Mg-rich (blue), and
Fe-rich (green). As mentioned above, these three elements are convenient
proxies for crust-like, mantle-like and core-like material respectively. We
therefore also indicate with coloured crosses the abundance ratios of Earth's
continental-crust, mantle, and core. The interior region bounded by these three
points represents Ca/Mg/Fe values that can be decomposed into crust/mantle/core
fractions assuming Earth-like compositions for each structural layer.

It is apparent from the diagram that such a decomposition is impossible for
almost half of the DZ white dwarfs in our sample as they lie exterior to the
triangle formed by the Earth's crust/mantle/core points. However, it is crucial
to notice that most of these are clustered towards the Mg-rich corner
(particular for values of $\mathrm{Fe}/[\mathrm{15Ca} + \mathrm{Mg} +
\mathrm{Fe}] < 0.3$). We take this as evidence for Mg enhancement related to
relative diffusion as discussed in Section~\ref{reldiff}.

It is therefore clear that for most DZ white dwarfs, the effects of relative
diffusion have to be considered when discussing the parent body compositions.
The exception to this rule is when the Mg fraction is particularly low, i.e.
the extremely  Ca-rich and Fe-rich systems highlighted in red and green
respectively in Fig.~\ref{fig:ternary}.

Ignoring Mg for a moment, we are able to make some statements about the
distribution of exoplanetesimal compositions by considering only the Fe to Ca
ratio. As described in Section~\ref{reldiff}, our envelope calculations show Ca
and Fe to have similar diffusion time-scales. Therefore over the few (no more
than about 10) diffusion time-scales that the material can remain visible, the
ratio of Fe to Ca changes only by a small amount compared to our measurement
errors.

\begin{figure}
  \centering
  \includegraphics[angle=00,width=\columnwidth]{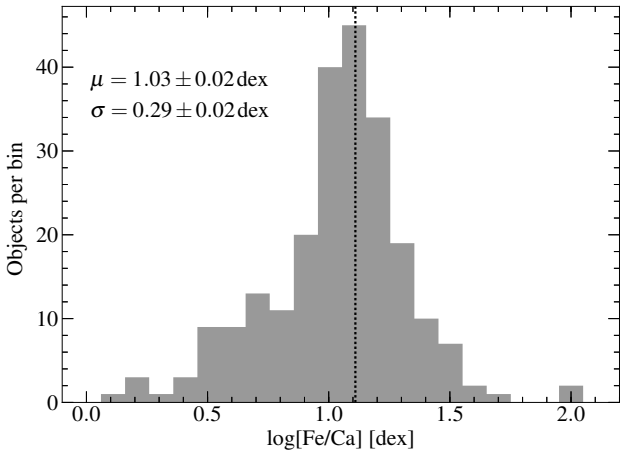}
  \caption{\label{fig:feca}
  The distribution of \logXY{Fa}{Ca} for our DZ sample, with the bulk Earth
  value indicated by the vertical dashed line. The mean and standard deviation
  of the sample are given by $\mu$ and $\sigma$ respectively.}
\end{figure}

The distribution of \logXY{Fe}{Ca} spans two orders of magnitude
(Fig.\,\ref{fig:feca}), and is approximately Gaussian in shape, although with a
possible excess of systems in the low Fe/Ca wing. The mean and standard
deviation of the distribution are found to be $1.03\pm0.02$\,dex and
$0.29\pm0.02$\,dex respectively, where the errors are estimated from
bootstrapping the data.

While it is tempting to interpret this distribution as directly representative
of exoplanetesimal compositions, this is unlikely to be true. The wings of the
distribution, where the Fe/Ca ratio is most extreme, indicate white dwarfs
which have almost certainly accreted material from a single large parent body
with an extreme composition. The atmospheric Fe/Ca ratios of white dwarfs
located near the centre of the distribution can be explained through several
different accretion histories. While accretion of single large asteroids with
$\logXY{Fe}{Ca} \approx 1.1$\,dex could be true for some of the stars, other
possibilities include accretion of multiple smaller asteroids with a bulk Earth
composition, or even multiple planetesimals with wildly different compositions
which average to a near bulk Earth Fe/Ca value. This means that our observed
distribution must be narrower than the true distribution of planetesimal
compositions. How much narrower depends not only on the underlying
\logXY{Fe}{Ca} distribution but also on the planetesimal mass distribution
(which may itself be a function of \logXY{Fe}{Ca}), which will determine how
the averaging of compositions from multiple planetesimals is statistically
weighted. Nevertheless, the distribution we derive is likely to have a mean
close to that of the underlying $\logXY{Fe}{Ca}$ distribution, which
Fig.~\ref{fig:feca} suggests is close to the bulk Earth ratio Furthermore, the
distribution in Fig.~\ref{fig:feca} highlights the sample size required to
detect systems with $\logXY{Fe}{Ca}$ ratios of $\pm 1$\,dex from the mean.

\subsection{Structural interpretation of planetary material}
\label{structure}

In Fig.~\ref{fig:ternary}, we have highlighted several white dwarfs that have
accreted material with atypically high Ca- and Fe-rich compositions, which we
have qualitatively described as being crust- and core-like, respectively. These
descriptions are justified by the proximity to the locations of the Earth's
crust and core within Fig.~\ref{fig:ternary}. In Fig.~\ref{fig:feca} these same
objects are found in the bins at both ends of the distribution, spanning two
orders of magnitude in their Fe to Ca ratios.

Here we demonstrate that we can be far more quantitative in assessing the
crust- and core-like nature of the accreted Ca- and Fe-rich material if we make
the assumptions that (1) that rocky planetesimals can generally be described as
a mixture, i.e. linear combination, of crust, mantle and core, and (2) the
abundances of the Earth's crust/mantle/core are typical for differentiated
planet(esimals). The first assumption is not strictly true if applied to
primitive planetesimals, i.e. chondrites, which are homogenous in their
composition, lacking the distinct geological layers that result from
differentiation. Instead we can consider chondrites as a linear combination of
potential crust/mantle/core had they undergone differentiation. The second
assumption appears to be justified as none of the stars we analysed indicate
planetary debris more Ca-rich than the Earth's crust nor more Fe-rich than the
Earth's core (Fig.~\ref{fig:ternary}).

To apply the above reasoning to the white dwarfs analysed in our sample, we
first consider an exoplanetesimal whose total mass is the linear sum of its
crust, mantle, and core components, denoted as $M_\mathrm{Cru}$,
$M_\mathrm{Man}$, and $M_\mathrm{Cor}$, respectively (assumption 1). Applying
assumption 2, that the atomic compositions of these three geological components
are identical to those of the Earth, we then know each of their mass-fractions
of Ca, Mg, and Fe. For instance 5.2 percent of Earth's crust is comprised of
Fe, which can be written as a matrix element $\mathrm{Cru_{Fe}} = 0.052$. Thus
the elemental masses $M_\mathrm{Ca}$, $M_\mathrm{Mg}$, and $M_\mathrm{Fe}$ for
the entire asteroid are calculated via

\begin{equation}
  \begin{pmatrix}
  M_\mathrm{Ca} \\
  M_\mathrm{Mg} \\
  M_\mathrm{Fe}
  \end{pmatrix}
  = \begin{pmatrix}
  \mathrm{Cru_{Ca}} & \mathrm{Man_{Ca}} & \mathrm{Cor_{Ca}} \\
  \mathrm{Cru_{Mg}} & \mathrm{Man_{Mg}} & \mathrm{Cor_{Mg}} \\
  \mathrm{Cru_{Fe}} & \mathrm{Man_{Fe}} & \mathrm{Cor_{Fe}}
  \end{pmatrix} \times \begin{pmatrix}
  M_\mathrm{Cru} \\
  M_\mathrm{Man} \\
  M_\mathrm{Cor}
  \end{pmatrix},
  \label{eq:matrices1}
\end{equation}
where, using values obtained from \citet{rudnick+gao03-1},
\citet{palme+oneill03-1}, and \citet{mcdonough00-1} for the Earth's bulk
continental crust, mantle, and core respectively, the complete set of matrix
elements are 
\begin{equation}
  \begin{pmatrix}
  \mathrm{Cru_{Ca}} & \mathrm{Man_{Ca}} & \mathrm{Cor_{Ca}} \\
  \mathrm{Cru_{Mg}} & \mathrm{Man_{Mg}} & \mathrm{Cor_{Mg}} \\
  \mathrm{Cru_{Fe}} & \mathrm{Man_{Fe}} & \mathrm{Cor_{Fe}}
  \end{pmatrix} = \begin{pmatrix}
   0.046 & 0.026 & 0 \\
   0.028 & 0.222 & 0 \\
   0.052 & 0.063 & 0.85 \end{pmatrix}.
  \label{eq:matrices2}
\end{equation}

For the interpretation of the debris abundances determined from the
spectroscopic analysis of our white dwarf sample, we approach the problem from
the opposite direction, i.e. having measured Ca, Mg, and Fe abundances but
wishing to determine the relative contributions of the three structural
components. Since \eqref{eq:matrices2} is a non-degenerate square matrix, this
is easily achieved through inversion of equation\,\eqref{eq:matrices1}. Note
that while equation\, \eqref{eq:matrices1} is defined in terms of masses, mass
accretion rates (which are commonly found in the literature for systems assumed
to be in accretion/diffusion equilibrium) can equally be used in their place.
Since our goal is to determine structural component \emph{fractions}, this
simply means normalising such that they sum to one, and so no specific
normalisation of Ca/Mg/Fe is required beforehand. A point worth mentioning is
that so far we have been referring to atomic abundances, which is the standard
convention in atomic spectroscopy. Here we require mass abundances, and so each
of the Ca/Mg/Fe atomic abundances requires rescaling by its atomic mass.

\begin{table}
  \begin{center}
    \caption{\label{tab:struct} 
    Estimated crust/mantle/core mass-fractions for the objects shown in
    Fig.~\ref{fig:ternary2}, and calculated according to
    equation~\eqref{eq:matrices1}. The first set of white dwarfs are the
    Ca-rich and Fe-rich systems from our sample. This is compared below with
    objects from the literature where accretion rates for Ca, Mg, and Fe are
    available.
    }
  \begin{tabular}{lrrrc}
  \hline
  Name             & Crust & Mantle & Core & Ref. \\
  \hline
  SDSS\,J0741+3146  & $ 0.080$ & $ 0.043$ & $ 0.877$ & 1     \\
  SDSS\,J0744+4649  & $ 0.798$ & $ 0.158$ & $ 0.044$ & 1     \\
  SDSS\,J0823+0546  & $ 0.067$ & $ 0.091$ & $ 0.843$ & 1     \\
  SDSS\,J1033+1809  & $ 0.897$ & $-0.058$ & $ 0.161$ & 1     \\
  SDSS\,J1043+3516  & $ 0.102$ & $ 0.179$ & $ 0.719$ & 1     \\
  SDSS\,J1055+3725  & $ 0.881$ & $ 0.001$ & $ 0.118$ & 1     \\
  SDSS\,J1351+2645  & $ 0.897$ & $-0.058$ & $ 0.161$ & 1     \\
  \hline
  GD\,16            & $ 0.273$ & $ 0.420$ & $ 0.307$ & 2     \\
  GD\,17            & $ 0.302$ & $ 0.415$ & $ 0.283$ & 2     \\
  GD\,40            & $ 0.770$ & $ 0.132$ & $ 0.098$ & 3,4   \\
  GD\,61            & $ 0.384$ & $ 0.584$ & $ 0.032$ & 5     \\
  SDSS\,J0738+1835  & $-0.169$ & $ 0.829$ & $ 0.340$ & 6     \\
  NLTT\,19868       & $ 0.777$ & $ 0.248$ & $-0.025$ & 7     \\
  SDSS\,J0845+2257  & $ 0.004$ & $ 0.497$ & $ 0.499$ & 8     \\
  PG\,1015+161      & $ 0.309$ & $ 0.325$ & $ 0.366$ & 9     \\
  SDSS\,J1043+0855  & $ 0.618$ & $ 0.383$ & $-0.001$ & 10    \\
  NLTT\,25792       & $ 0.183$ & $ 0.600$ & $ 0.218$ & 11    \\
  WD\,1145+017      & $ 0.188$ & $ 0.355$ & $ 0.457$ & 12    \\
  PG\,1225$-$079    & $ 0.763$ & $ 0.047$ & $ 0.190$ & 4     \\
  SDSS\,J1228+1040  & $ 0.480$ & $ 0.369$ & $ 0.150$ & 9     \\
  SDSS\,J1242+5226  & $ 0.307$ & $ 0.570$ & $ 0.123$ & 13    \\
  G\,149$-$28       & $ 0.356$ & $ 0.444$ & $ 0.200$ & 14    \\
  WD\,1536+520      & $-0.016$ & $ 0.782$ & $ 0.234$ & 15    \\
  NLTT\,43806       & $ 0.579$ & $ 0.380$ & $ 0.042$ & 14    \\
  GD\,362           & $ 0.850$ & $-0.022$ & $ 0.172$ & 4     \\
  GALEX\,J1931+0117 & $-0.191$ & $ 0.578$ & $ 0.613$ & 16,17 \\
  G\,241$-$6        & $ 0.520$ & $ 0.384$ & $ 0.096$ & 3,4   \\ %Also Zuckerman(2011)
  HS\,2253+8023     & $ 0.374$ & $ 0.327$ & $ 0.299$ & 18    \\
  G\,29$-$38        & $ 0.448$ & $ 0.376$ & $ 0.175$ & 19    \\
  \hline
\end{tabular}
  \end{center}
  \begin{flushleft}
  References:
  (1)~This work/Paper~I,
  (2)~\citet{gentileetal17-1},
  (3)~\citet{juraetal12-1},
  (4)~\citet{xuetal13-1},
  (5)~\citet{farihietal13-1},
  (6)~\citet{dufouretal12-1},
  (7)~\citet{kawka+vennes16-1},
  (8)~\citet{wilsonetal15-1},
  (9)~\citet{gaensickeetal12-1},
  (10)~\citet{melis+dufour17-1},
  (11)~\citet{vennes+kawka13-1},
  (12)~\citet{xuetal16-1},
  (13)~\citet{raddietal15-1},
  (14)~\citet{zuckermanetal11-1},
  (15)~\citet{farihietal16-1},
  (16)~\citet{vennesetal11-1},
  (17)~\citet{melisetal11-1},
  (18)~\citet{kleinetal11-1},
  (19)~\citet{xuetal14-1}.
  \end{flushleft}
\end{table}

\begin{figure*}
  \centering
  \includegraphics[angle=00,width=0.8\textwidth]{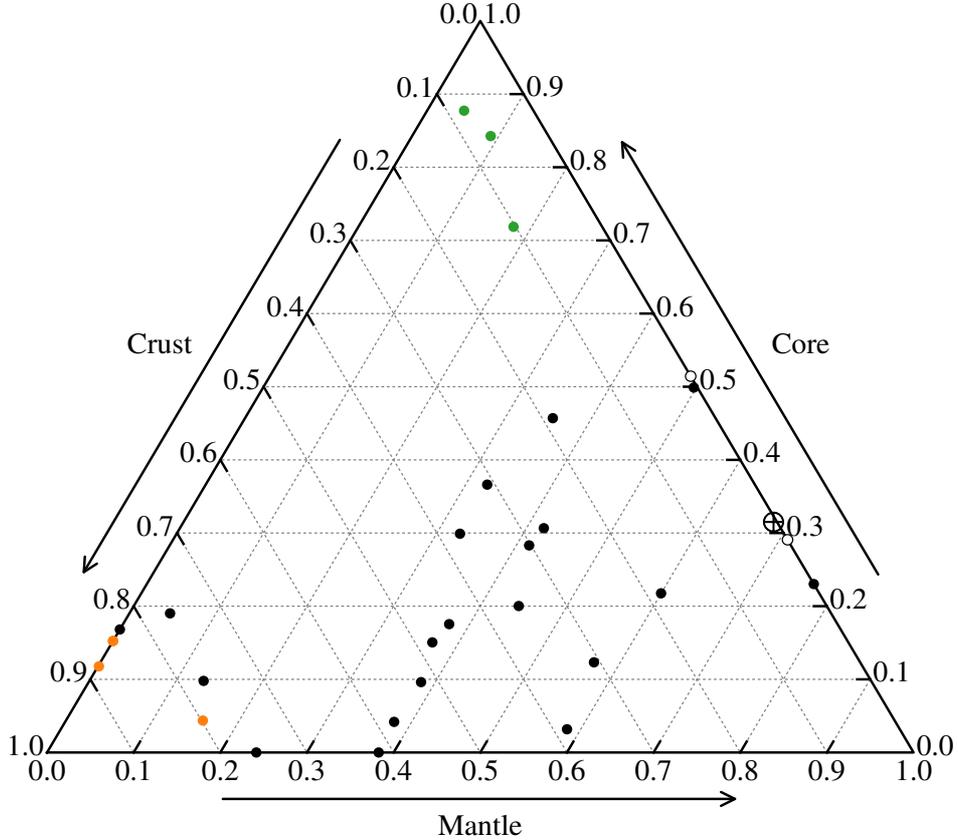}
  \caption{\label{fig:ternary2}
  Ternary diagram illustrating the mass fraction of accreted planetary debris
  in terms of crust, mantle, and core material. The outliers in our sample are
  shown in orange/green for Ca/Fe rich white dwarfs respectively (discussed at
  length in Section~\ref{extreme}). White dwarfs from other published abundance
  studies are shown in black (Table~\ref{tab:struct}), and as open circles for
  two cases where the calculated crust value is $\simeq -0.2$ and so
  significant clipping was required to move these points to the boundary of the
  diagram. The bulk Earth is indicated by the $\oplus$ symbol. 
  }
\end{figure*}

Because of the effects of relative diffusion on Mg abundances
(Section~\ref{reldiff}), we restrict our application of this approach to the
seven Ca- and Fe-rich systems in our sample (Fig.~\ref{fig:ternary}) where the
calculated crust/mantle/core values are largely insensitive to the low Mg
fractions. For the other white dwarfs where abundances are strongly affected by
relative diffusion, application of equation\,\eqref{eq:matrices1} yields
nonsensical negative crust fractions and/or and mantle fractions exceeding 100
percent. We instead complement these seven Ca- and Fe-rich DZ white dwarfs with
detailed abundance studies of 22 warmer and younger debris-polluted white
dwarfs which are often assumed to be in accretion-diffusion equilibrium. This
means their atmospheric abundances can be corrected for the different sinking
times of the individual elements and hence the bulk compositions of the parent
bodies from the accretion rate of each element. We present the results of our
structural decomposition in Table~\ref{tab:struct} and graphically in
Fig.~\ref{fig:ternary2}. The additional systems were all selected with the
requirement that all of Ca, Mg, and Fe had spectroscopic detections with
calculated accretion rates. Therefore potentially interesting white dwarfs such
as PG\,0843+225 \citep{gaensickeetal12-1}, and GD\,133 \citep{xuetal14-1} could
not be included at this time, as one of Ca/Mg/Fe had only an upper limit
available. We also exclude WD\,1425+540 \citep{xuetal17-1} as it would be
senseless to try and explain cometary material in terms of rocky geology.

For eight systems, application of equation \eqref{eq:matrices1} resulted in a
negative value for one structural component. For display purposes, we set the
negative values to 0, and renormalised the other two components such that they
appear on the boundary of Fig.~\ref{fig:ternary2}. For six of these the effect
is only minor (only a few $0.01$ or less), but for \sdss{0738}{+}{1835} and
GALEX\,J1931$+$0117 the calculated crust values are $\simeq-0.2$ (displayed in
grey in Fig.~\ref{fig:ternary2}). This issue can be resolved if the parent body
mantles were relatively Ca-poor compared to the Earth's, suggesting our second
assumption (above) is not universally applicable. We find that reducing
$\mathrm{Man_{Ca}}$ in equation~\eqref{eq:matrices2} by a factor of 2.5 is
sufficient to move both \sdss{0738}{+}{183} and GALEX\,J1931+0117 within the
bounds of Fig.~\ref{fig:ternary2}, and thus with compositions consistent with
combinations of mantle and core material. Additionally, it is suggested by
\citet{dufouretal12-1} for \sdss{0738}{+}{183} and \citet{melisetal11-1} for
GALEX\,J1931+0117, that their compositions may be indicative of stripping of
their outer layers, including part of its mantle for GALEX\,J1931+0117.

We find the majority of objects from published analyses (black points in
Fig.~\ref{fig:ternary2}) are consistent with crust-like material or a mixture
of crust and mantle rather than the bulk Earth (although, we note that
WD\,1536$+$520 is only $\simeq0.1$ fractional plot units away). For instance
the accreted material at NLTT\,43806 is described as being in best agreement
with ``a mixture of terrestrial crust and upper mantle material'' by
\citet{zuckermanetal11-1}, whose analysis also included the detection of Al
accretion. The crust/mantle/core mass fractions we calculated (Table
\,\ref{tab:struct}) corroborate this assertion. Similarly, we find that among
the previously published analyses (Table~\ref{tab:struct}),
\sdss{0845}{+}{2257} has the highest core mass fraction, in good agreement with
the more detailed study of \citet{wilsonetal15-1} who detected large abundances
of Fe and Ni. Furthermore, \citet{wilsonetal15-1} argued for a mantle/core
mixture, but mantle depleted with respect to the Earth. We reach the same
conclusion from our analysis, where we find a mantle/core mixture of 50/50 for
\sdss{0845}{+}{2257} versus 70/30 for the Earth. These comparisons demonstrate
the effectiveness of our relatively simple approach to classifying
exoplanetesimal compositions, which only requires abundance measurements of Ca,
Mg, and Fe.

The overall banana-shaped distribution in Fig.~\ref{fig:ternary2} may initially
come as a surprise, however it is simple to see that this is indeed the
expected distribution of crust, mantle, core combinations. The dearth of points
along the right edge of the plot corresponds to the absence of parent bodies
made of a crust+core mixture but lacking a mantle, which is consistent with the
expectation that planetary objects undergoing differentiation will form with a
significant mantle component, in addition to their core and crust. Finally
there are no points in Fig.~\ref{fig:ternary2} corresponding to more than 80
percent mantle, whereas multiple objects are seen with more than 80 percent
crust or core compositions. While it may be possible to create mantle-dominated
asteroids via stripping of a larger body, the absence of points in this region
potentially indicates that such a process rarely occurs.

The distribution of points in Fig.~\ref{fig:ternary2} makes it clear that rocky
material accreted by white dwarfs often originates from highly differentiated
parent bodies. In particular, the prevalence of crust-dominated and
crust+mantle points suggests parent bodies originating from collisional
fragments of the upper layers of (minor-)planets as proposed by
\citet{zuckermanetal11-1} for NLTT\,43806. This argument becomes especially
appealing on consideration that the Earth's crust contributes less than
1~percent of its mass, yet for most of the objects in
Fig.~\ref{fig:ternary2}/Table~\ref{tab:struct} the associated crust fractions
are in range 15--90 percent.

\section{Extreme abundance ratios}
\label{extreme}

Instantaneous accretion can not be assumed for DZ white dwarfs due to the long
time-scales for metals to diffuse out of the base of their CVZs
(Section~\ref{reldiff}). Therefore in the general case, it is not possible to
establish whether the observed metal contamination arises from the accretion of
a single large object, or from multiple accretion episodes involving smaller
parent bodies. However several systems in our sample show compositions
consistent with the accretion of highly differentiated parent bodies in
particular those that are rich in Ca or Fe as discussed in Section~
\ref{composition}. These white dwarfs have very likely accreted single large
parent bodies because subsequent accretion episodes of many small planetesimals
(that were not previously part of a single larger object) are  expected to
average out to a less extreme abundance pattern, e.g. more similar to the bulk
Earth.

\begin{figure*}
  \centering
  \includegraphics[angle=00,width=0.8\textwidth]{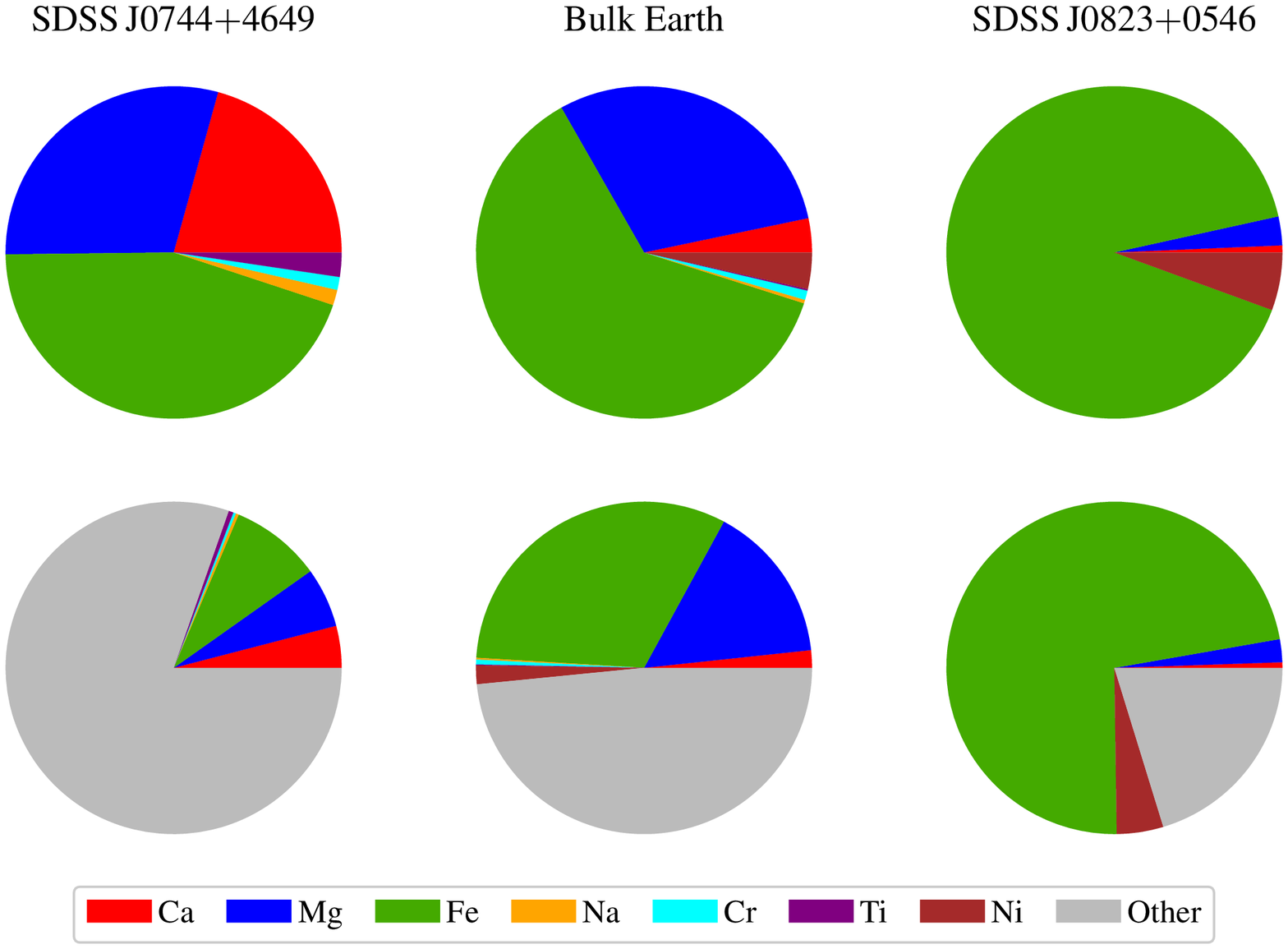}
  \caption{\label{fig:pies}
  The mass fractions of the accreted material for two of the most extreme DZ
  white dwarfs in our sample compared with the bulk Earth. The top row includes
  only the elements we detected in the observed spectra. In the bottom row, we
  use equation\,\eqref{eq:matrices1} to estimate the fraction of undetected
  material (which is likely to be dominated by O and Si), which we simply label
  as ``other''. The material accreted by \sdss{0744}{+}{4649} has a composition
  consistent with pure crust, whereas core material is implied for the parent
  body accreted by \sdss{0823}{+}{0546}.
  }
\end{figure*}

\subsection{Ca-rich objects}
\label{ca_rich}

\begin{figure*}
  \centering
  \includegraphics[angle=0,width=\textwidth]{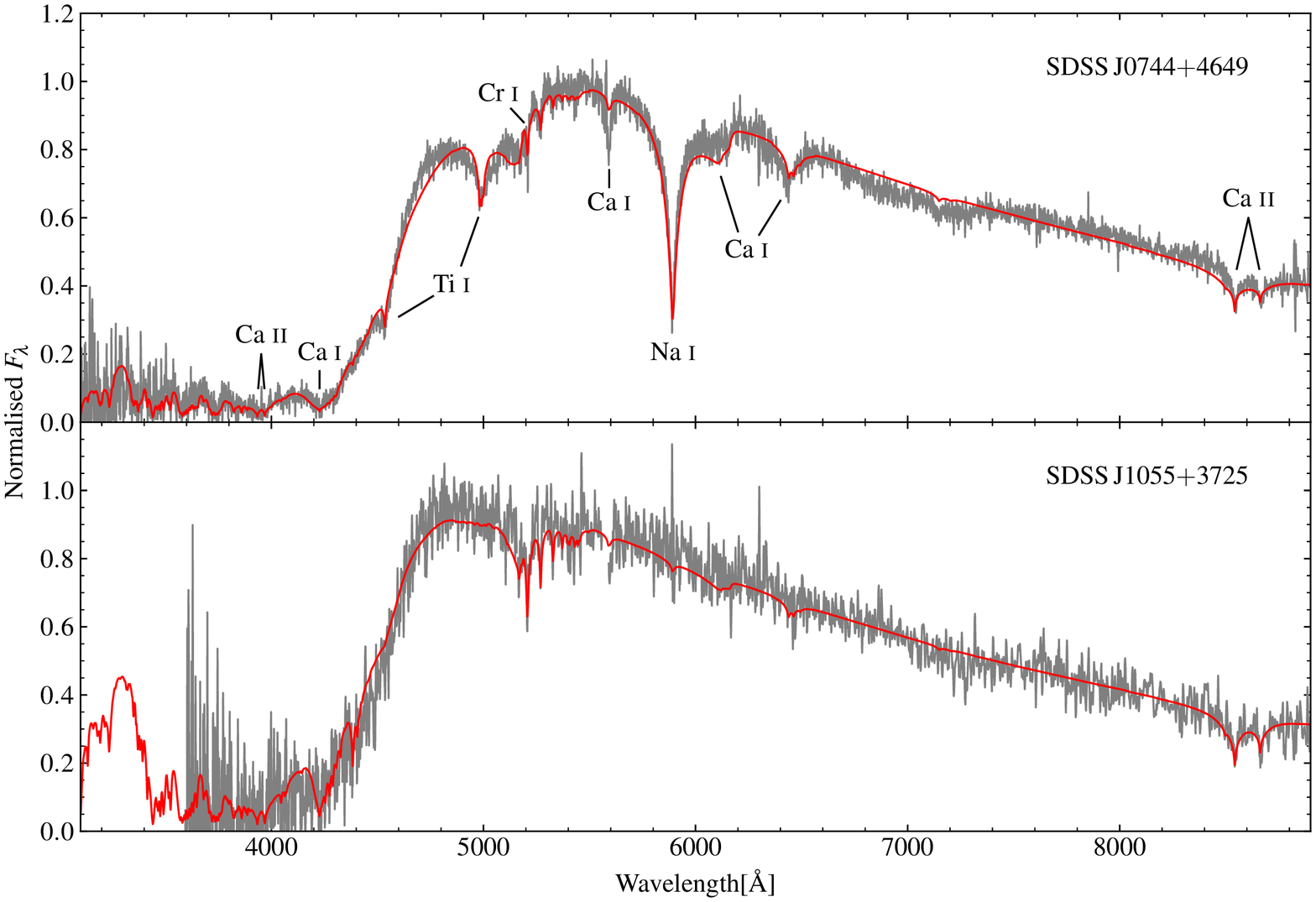}
  \caption{\label{fig:spectra_Ca}
  Two DZs that have accreted Ca-rich parent bodies. The two other Ca-rich DZs
  (\sdss{1033}{+}{1809} and \sdss{1351}{+}{2645}) are spectroscopically similar
  to \sdss{1055}{+}{3725}, and are thus not shown here (see the appendix of
  Paper~I for the spectra of the entire DZ sample).
  }
\end{figure*}

The four Ca-rich DZ white dwarfs we have identified are \sdss{0744}{+}{4649},
\sdss{1033}{+}{1809}, \sdss{1055}{+}{3725}, and \sdss{1351}{+}{2645}\footnote{
\sdss{1033}{+}{1809} and \sdss{1351}{+}{2645}, have degenerate
crust/mantle/core values and so are indistinguishable in
Fig.~\ref{fig:ternary2}.}, which are shown by the red points in
Fig.~\ref{fig:ternary}/\ref{fig:ternary2}. These are contenders for the most
crust-like in nature with similar values to GD\,40
\citep{juraetal12-1,xuetal13-1}, NLTT\,19868 \citep{kawka+vennes16-1},
PG\,1225$-$079 \citep{xuetal13-1}, and GD\,362 \citep{xuetal13-1}. The most
striking spectral feature of these DZs (Fig.~\ref{fig:spectra_Ca}) is their
huge increase in opacity bluewards of $\simeq4500$\,\AA, resulting in the
suppression of flux at shorter wavelengths. This arises from the extremely
pressure-broadened wings of the \Ion{Ca}{ii} H+K lines.

The most noteworthy of the Ca-rich systems is \sdss{0744}{+}{4649}. The
spectrum of this white dwarf is unique with no similar looking stars known.
Compared to calcium, both magnesium and iron are depleted with Fe/Ca and Mg/Ca
ratios $\simeq0.9$\,dex lower than those of the bulk Earth. The spectrum of
this star also exhibits strong lines of Ti and Na which are typically not seen
for other DZs with $\Teff\simeq 5000$\,K. The abundance ratios are
\logXY{Ti}{Ca}$ = -1.02$\,dex and \logXY{Na}{Ca}$ = -0.90$\,dex, where the
respective abundances are $-1.4$ and $-0.7$\,dex for the bulk Earth
\citep{mcdonough00-1}, and $-1.1$ and $-0.1$\,dex for the Earth's crust
\citep{rudnick+gao03-1}. The Ti/Ca ratio for \sdss{0744}{+}{4649} is much
closer to the Earth's crust value than the bulk Earth ratio, and as both
elements are refractory lithophiles, this reinforces the crust-like
interpretation of the accreted planetesimal. On the other hand, the Na/Ca ratio
is more similar to the bulk Earth ratio. However we note that while Na is a
lithophile, it is also a volatile element and so is not expected to condense at
the same temperature as Ca or Ti. Like Ca and Ti, Al is also a refractory
lithophile and thus is likely to be present at an abundance of $\logXY{Al}{Ca}
\simeq +0.4$\,dex \citep{rudnick+gao03-1}. However, the only strong optical
\Ion{Al}{i} transitions are located at 3944\,\AA\ and 3961\,\AA, between the
already saturated Ca H+K lines. As there is so little emergent flux in this
wavelength region (Fig.~\ref{fig:spectra_Ca}) detecting Al at
\sdss{0744}{+}{4649} would be extremely challenging even with improved
instrumentation. We also detect the moderately refractory lithophile Cr at
relative abundance of $\logXY{Cr}{Ca} = -1.3$\,dex, much higher than the trace
$-2.6$\,dex for the Earth's crust \citep{rudnick+gao03-1}. The mass fractions
for each element including presumed unseen elements are demonstrated by the
left-hand pie charts in Fig.~\ref{fig:pies}. If the material accreted by
\sdss{0744}{+}{4649} is indeed lithospheric, then Fig.~\ref{fig:pies} indicates
that the unseen elements make up 80\,percent of the total mass of the parent
body, with most of this comprised of O, followed by Si and then Al
\citep{rudnick+gao03-1}. Note that the optical transitions of O and Si become
extremely weak for the \Teff\ range of our sample, thus prohibiting their
detection in DZ white dwarfs.

For the other three Ca-rich objects, we are not able to offer an analysis as
detailed as for \sdss{0744}{+}{4649}, however they are worthy of discussion
none the less. All three stars have qualitatively similar spectra, with
\sdss{1055}{+}{3725} shown in Fig.~\ref{fig:spectra_Ca} (spectra for all DZs in
our sample are shown in Paper~I). As their SDSS spectra are of much lower in
quality than that of \sdss{0744}{+}{4649}, Ti is only detected for
\sdss{1351}{+}{2645} at a relative abundance of $\logXY{Ti}{Ca} \simeq
-0.6$\,dex, again supporting a crust-like interpretation of the accreted parent
body. At both \sdss{1055}{+}{3725} and \sdss{1351}{+}{2645} we also detect Cr
at relative abundances of $-1.0$ and $-1.2$\,dex respectively. In all three
detections of Cr (including \sdss{0744}{+}{4649}), the Cr/Ca ratios are found
to be greatly enhanced relative to the Earth's crust ($-2.6$\,dex), with
potential implications for their planetary formation conditions.
\citet{mcdonough00-1} notes that while regarded as a lithophile, under high
pressure, Cr exhibits siderophile behaviour, and thus for the Earth is
concentrated into the core \citep{moynieretal11-1}, where the bulk Earth value
is $\logXY{Cr}{Ca} = -0.7$\,dex. Therefore the parent bodies accreted by
\sdss{0744}{+}{4649}, \sdss{1055}{+}{3725}, and \sdss{1351}{+}{2645} were
likely of much lower mass than the Earth and thus formed under lower pressure
conditions where Cr exhibits lithophile behaviour.

\subsection{Fe-rich objects}
\label{fe_rich}

\begin{figure*}
  \centering
  \includegraphics[angle=0,width=\textwidth]{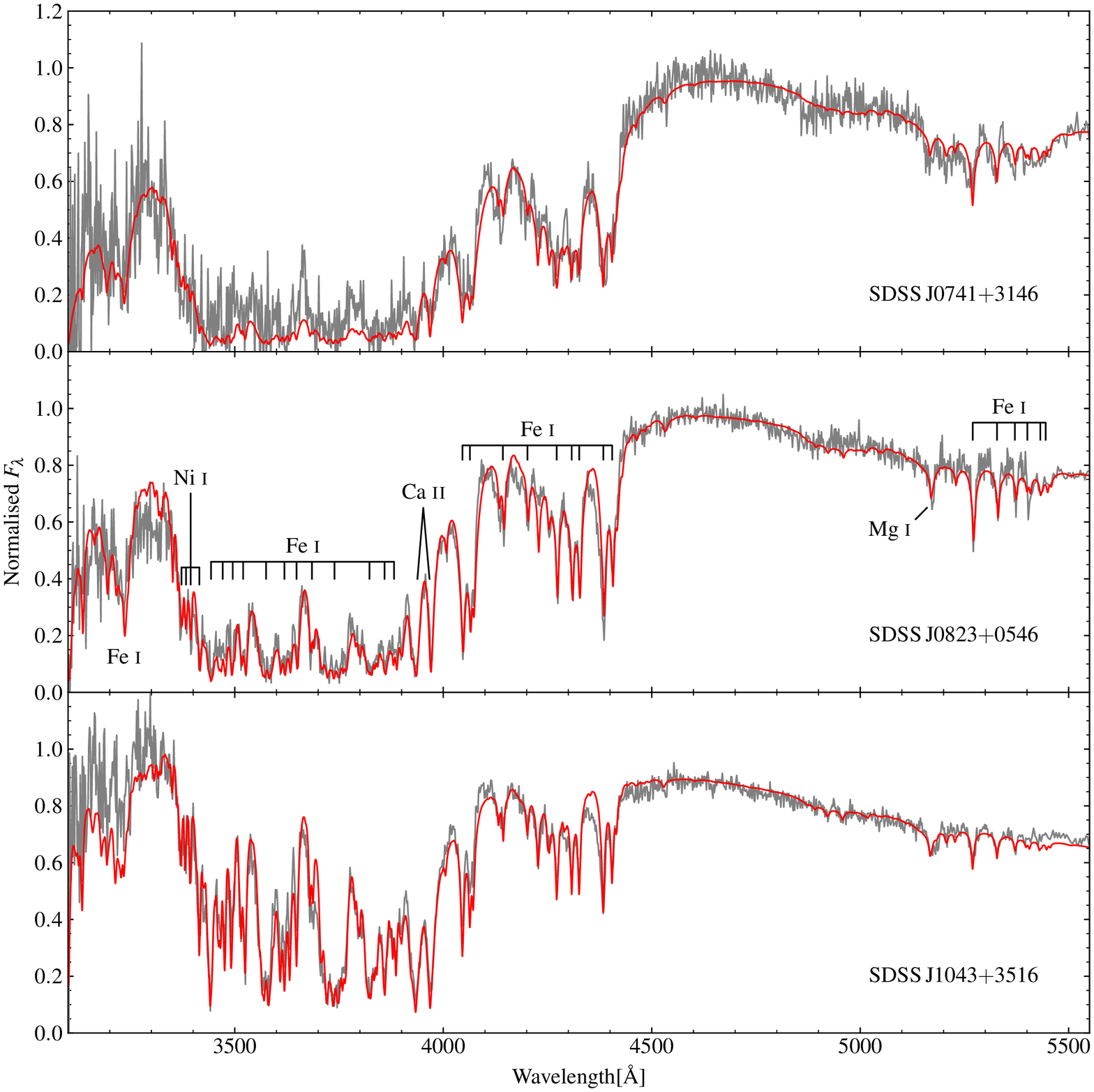}
  \caption{\label{fig:spectra_Fe}
  WHT spectra for the three objects which have accreted Fe-rich material.
  Because the WHT spectra extend as far blue as 3100\,\AA, we are able to
  constrain Ni abundances from the blend of lines at 3390\,\AA. The labelled
  spectrum of \sdss{0823}{+}{0546} shows that almost all absorption in these
  spectra comes from \Ion{Fe}{i} transitions, especially in the $3400$--$3900$
  range.
  }
\end{figure*}

The inferred compositions for the material accreted by \sdss{0741}{+}{3146} and
\sdss{0823}{+}{0546} are both extremely Fe-rich with
$\logXY{Fe}{Ca}>1.9$\,dex, the highest ratio known for any metal polluted white
dwarfs (see \citet{gaensickeetal12-1}, \citet{kawka+vennes16-1} and
\citet{wilsonetal15-1} for additional Fe-rich systems). The spectra of both
stars are quite similar (Fig.~\ref{fig:spectra_Fe}) showing a dense forest of
blended \Ion{Fe}{i} lines in the range 3400--3900\,\AA. Other notable Fe
features include the strong \transition{3}{F}{5}{G} triplet near 4400\,\AA, and
the \transition{5}{F}{5}{D} multiplet between 5250--5500\,\AA\ which provide
additional constraints on the Fe abundance and \Teff. While Fe is the dominant
contaminant in the atmospheres of these two stars, the intrinsic strengths of
the Ca H+K resonance lines result in the low Ca abundances remaining well
constrained. \sdss{1043}{+}{3516}, is the next most Fe-rich in our sample with
\logXY{Fe}{Ca} = 1.68\,dex. Its spectrum is qualitatively similar to those of
\sdss{0741}{+}{3146} and \sdss{0823}{+}{0546}, however the Fe lines are
slightly weaker and H+K lines slightly stronger.

From our structural analysis in Section~\ref{structure}, \sdss{0741}{+}{3146}
and \sdss{0823}{+}{0546} appear to be the most core-like exoplanetesimals
discovered to date. We note that the Fe-rich system NLTT\,888 comes close with
\logXY{Fe}{Ca} of 1.76\,dex \citep{kawka+vennes14-1,kawka+vennes16-1}, however
the lack of firm Mg measurement precludes it from being placed on
Fig.~\ref{fig:ternary2}, i.e. using either the quoted upper limit or precisely
0 for the Mg abundance yield wildly different results for the crust/mantle/core
decomposition. Using the upper limit of $\logX{Mg} <-8.7$\,dex leads to a
negative crust-value, however by setting \logX{Mg} to $-9.6$\,dex places
NLTT\,888 at the edge of Fig.~\ref{fig:ternary2} with a mantle fraction of 30
percent and core fraction of 70 percent.

For all three Fe-rich systems we obtained spectra using the \emph{Intermediate
dispersion Spectrograph and Imaging System} (ISIS) on the William Herschel
Telescope (WHT). For \sdss{0823}{+}{0546}, which we identified as a DZ
candidate from its SDSS colours and its relatively high proper motion, this is
the only available spectrum. Details on the observations and reduction of the
WHT spectra are summarised in Paper~I. Because the WHT spectra extend as far
blue as 3100\,\AA, we are also able to constrain Ni abundances for all three
stars using a set of \Ion{Ni}{i} lines at $\simeq 3390$\,\AA, and measure
\logXY{Fe}{Ni} as 1.59, 1.23, and 1.35\,dex for \sdss{0741}{+}{3146},
\sdss{0823}{+}{0546}, and \sdss{1043}{+}{3516} respectively\footnote{Note that
for \sdss{0741}{+}{3146}, the relative flux errors around 3400\,\AA\ are quite
large and a weak magnetic field of $0.48\pm0.05$\,MG appears to be present
(identified after publication of Paper~I). Therefore, the Ni abundance of
\sdss{0741}{+}{3146} is probably not as well constrained as the other two
systems.}. For the bulk Earth and the Earth's core, \logXY{Fe}{Ni} is about
1.24 and 1.20\,dex respectively \citep{mcdonough00-1} which is particularly
close to the 1.23\,dex measured for \sdss{0823}{+}{0546} -- the most precisely
analysed Fe-rich object in our sample due to the exceptional quality of the WHT
spectrum (Fig.~\ref{fig:spectra_Fe}). We note that these Fe/Ni ratios are
consistent with the metallic alloy kamacite which is predominantly Fe with a
few percent Ni, and is found in metallic meteorites. In contrast
\citet{wilsonetal15-1} find a much higher Ni content consistent with taenite or
a taenite/kamacite mixture.

Our structural decomposition of these systems combined with the Fe/Ni ratios
leave little ambiguity that the accreted planetesimals underwent
differentiation during their formation. The differentiated bodies must then
have undergone stripping of their crust and mantle, resulting in
exoplanetesimals comprised primarily of core material, which were subsequently
accreted on to these white dwarfs.

For both \sdss{0741}{+}{3146} and \sdss{1043}{+}{3516} Cr is also detected via
the 5207\,\AA\ \Ion{Cr}{i} line, but is notably absent from the spectrum of
\sdss{0823}{+}{0546} (despite the high signal-to-noise ratio of the spectrum).
Recalling that Cr is a generally regarded a lithophile but exhibits siderophile
behaviour at high pressure \citep{mcdonough00-1}, this may also suggest
different formation environments for the core material observed across these
systems. For \sdss{0741}{+}{3146} and \sdss{1043}{+}{3516} the atomic Cr/Fe
ratio is measured to be $-1.7$ and $-2.1$\,dex respectively, where
\citet{mcdonough00-1} quote a value of $-2.0$\,dex for the Earth's core.
Therefore we speculate that the parent bodies accreted by \sdss{0741}{+}{3146}
and \sdss{1043}{+}{3516} may have originated from planetary mass objects,
whereas for \sdss{0823}{+}{0546} the accreted planetesimal formed within a
lower mass body such as a minor planet.

Under the hypothesis that the Fe currently residing in the CVZ originates from
a single accretion episode, we can deduce a lower limit on the mass of the
parent body. In the following, we assume $\Mwd=0.6\,\Msun$, as the mass of the
CVZ depends on the white dwarf mass. This results in
$M_\mathrm{Fe}\simeq2\times10^{21}$\,g and
$M_\mathrm{Fe}\simeq3\times10^{21}$\,g for \sdss{0741}{+}{3146} and
\sdss{0823}{+}{0546}, respectively\footnote{Many other white dwarfs in our
sample exhibit larger abundances of Fe, but their unremarkable compositions are
consistent with the accumulation of material from multiple accretion episodes
or single large bodies with Earth-like compositions.}. These are, in fact,
conservative lower limits, as the planetary cores are not purely composed of
Fe, and as the debris composition suggests that the planetesimals were not
entirely made up of core-material. Furthermore some of the Fe/Ni may have
already sunk out of the base of convection zone depending on how long ago the
accretion events occurred. Application of equation\,\eqref{eq:matrices1} on the
crust/mantle/core fractions (Table~\ref{tab:struct}) implies Fe alone comprised
$\simeq 72$\,percent of the total parent body mass at \sdss{0823}{+}{0546}. If
we consider Ni as well as Fe, then these two core elements account for $\simeq
77$\,percent of the total accreted mass. Because the columns in
equation\,\ref{eq:matrices2} do not sum to one, the implication is that most of
the remaining mass includes undetected elements, such as O and Si
(Fig.~\ref{fig:pies}, bottom right). The measured compositions for
\sdss{0823}{+}{0546} and those implied by equation 4 are illustrated in the
right-hand pie charts of Fig.~\ref{fig:pies}. At a maximum density of
7.9\,\gcc\ (a pure Fe/Ni composition with no porosity), we arrive to minimum
geometric-mean radii of 39 and 45\,km for the planetesimals accreted by
\sdss{0741}{+}{3146} and \sdss{0823}{+}{0546}, respectively.

The above analysis assumes that the present CVZ metal masses correspond to the
total accreted masses. While these estimated masses are comparable to
moderately large Solar system asteroids, some of the material has presumably
already sunk out of the bases of their CVZs since the accretion events
occurred, and thus the parent bodies must have been larger. However, intuition
tells us that the accreted planetesimals can not have been much more massive,
as there ought to be fewer available bodies at higher mass-intervals.
Consequently the accretion episodes ought to have occurred within only a few
$\tau_\mathrm{Fe}$ ago. For example, if the material at \sdss{0823}{+}{0546}
was deposited $t=10\tau_\mathrm{Fe}$ ago, the implied parent body mass would be
similar to that of the Moon. While this is by no means impossible, a mass
closer to that observed now in the CVZ seems far more reasonable, thus implying
a more recent accretion history. We show here that this intuition can be
translated directly into statistics, providing median values and 95th
percentile upper-limits to the asteroid masses, and hence times since the
accretion episodes.

The most massive planetesimals in the Solar asteroid belt have a power-law
mass-distribution with an exponent $k \simeq 1.8$ \citep{kresak77-1}. We
therefore assume that the large large metallic exoplanetesimals accreted by
\sdss{0741}{+}{3146} and \sdss{0823}{+}{0546} have masses, $M$, drawn from a
similar distribution, $P(M)$, with the mass of material in the white dwarf CVZ,
$\Mcvz$ as a lower bound. $P(M)$ can then be written as
\begin{equation}
  P(M) = \frac{k-1}{\Mcvz} \left(\frac{M}{\Mcvz}\right)^{-k}\quad
  \mbox{for $M \geq \Mcvz$},\quad k > 1.
\label{eq:Mpdf}
\end{equation}
Integrating \eqref{eq:Mpdf} up to a mass $\Mast$, yields the corresponding
quantile $q$, i.e.
\begin{equation}
  q = \int_{\Mcvz}^{\Mast} P(M)\,\dd M
    = 1 - \left(\frac{\Mast}{\Mcvz}\right)^{1-k}.
\label{eq:quantile}
\end{equation}
Rearranging equation\,\eqref{eq:quantile} to express the $\Mast/\Mcvz$
ratio in terms of $q$,
\begin{equation}
  \Mast/\Mcvz = (1-q)^{1/1-k},
  \label{eq:mass_ratio}
\end{equation}
it then becomes simple to calculate the median and 95th percentile upper-limit
of $\Mast/\Mcvz$, for a given value of the power-law exponent, $k$. Using $k =
1.8$ as above we find $\Mast/\Mcvz$ has a median value of 2.4 and
95th-percentile upper-limit of 42. In other words, the initial planetesimal
masses were probably only a few times larger than what currently remains within
the CVZs, and is unlikely to be more than a few ten times larger. These upper
limits correspond to $\sim10^{23}$\,g or about one tenth the mass of Ceres, for
the metallic objects by \sdss{0741}{+}{3146} and \sdss{0823}{+}{0546}.

Continuing this line of reasoning, we can place similar constraints on how long
ago these planetesimals were accreted. Because their composition is dominated
by Fe and Ni (which both have similar diffusion time-scales), we are justified
in considering the time-evolution of material in the white dwarf CVZ as
\begin{equation}
  \Mcvz(t) = \Mast\mathrm{e}^{-t/\tau_\mathrm{Fe}}.
  \label{eq:dmdt}
\end{equation}
In the general case where the mass has components from elements with different
sinking time-scales, \eqref{eq:dmdt} becomes a sum of exponentials, which
cannot be analytically solved for $t$. Combining equations
\eqref{eq:mass_ratio} and \eqref{eq:dmdt}, we can then write the time since
accretion in terms of $q$ and $k$ as
\begin{equation}
  t = \tau_\mathrm{Fe}\frac{\ln(1-q)}{1-k}.
\end{equation}
Using again $k=1.8$, we find the median and 95-percentile upper-limits for $t$
are $0.87\,\tau_\mathrm{Fe}$ and $3.74\,\tau_\mathrm{Fe}$ respectively.

To place these systems in context, there are several comparable Solar system
objects that are worth consideration. This includes the main-belt asteroid,
16-Psyche, which is the largest M-type asteroid in the Solar system. Radar
observations show that Psyche has a mostly metallic composition
\citep{ostro85-1} consistent with exposed core material. At a mass of
$2.72\pm0.75\times10^{22}$\,g. \citep{carry12-1}, it is only one order of
magnitude larger than the estimated Fe mass currently residing in the CVZ of
\sdss{0823}{+}{0546}. While Psyche is thought to be chiefly comprised of Fe and
Ni \citep{matteretal13-1}, NIR observations indicate that its surface
composition also includes pyroxenes \citep{hardersen05-1}. As pyroxenes can
include Ca and Mg it is possible that these elements we see in these Fe-rich
white dwarfs could also arise from such compounds on the surface of the
metallic asteroids.

The metallic (as opposed to rocky) nature of these exoplanetesimals offers the
opportunity to investigate the process of the accretion on to the white dwarf
surface which we show to be violently destructive, and not necessarily leading
to the existence of a debris disc. Typically, planetesimals arriving at white
dwarfs are assumed to be loose rubble piles held together through
self-gravitation. Disruption occurs when tidal forces overcome self-gravity. In
the case of loose rubble piles, the distance from the white dwarf at which this
occurs (the Roche radius) depends only on the white dwarf mass and planetesimal
density, and is typically in the range 1--2\,\Rsun. For Fe-rich asteroids, the
mechanical strength of the material cannot be ignored, and a different
treatment is required. The effect of mechanical stresses on planetesimals has
previously been examined by some authors
\citep[e.g.][]{slyuta+voropaev97-1,davidsson99-1}, with \citet{brownetal17-1}
recently considering the disruption of high mechanical-strength planetesimals
in the context of white dwarf accretion. They show that for an asteroid with
density $\rho$, size $a_0$, and tensile strength $S$, a simple relation for the
tidal disruption radius $R_\mathrm{td}$ is given by
\begin{equation}
  R_\mathrm{td}^3 = \frac{GM_{\mathrm{wd}}\rho a_0^2}{2S},
  \label{eq:td}
\end{equation}
where $G$ is the gravitational constant and $M_\mathrm{wd}$ is the white dwarf
mass.

The small amount of literature available on the mechanical properties of
metallic Solar system bodies indicates their characteristics can vary
dramatically. For the meteorite samples that have been studied, tensile
strengths have been found ranging from 40\,MPa \citep{slyuta13-1} up to
800\,MPa \citep{opik58-1}, and exceeding 1\,GPa for some cast Fe-Ni alloys
\citep{petrovic01-1}. To explore an extreme example, we take $S=800$\,MPa.
Assuming the density for meteoritic-iron of 7.9\,\gcc, we estimated above
minimum radii of 39\,km for \sdss{0741}{+}{3146} and 45\,km for
\sdss{0823}{+}{0546}, respectively\footnote{The radii were likely larger than
this for two reasons. Firstly, as our above analysis shows, the masses could be
somewhat larger, and secondly if the asteroids had any significant porosity,
their effective densities would be lowered.}. Setting \Mwd\ to the canonical
0.6\,\Msun, equation\,\eqref{eq:td} implies similar tidal disruption distances
for both systems at $R_{\rm td} \simeq 0.17\,\Rsun = 13\,\Rwd$, much closer
than the $\simeq 1\,\Rsun$ for a strengthless rubble pile. Of course, if any
faults are present within the asteroid, this will allow for disintegration at
greater distances from the white dwarf, but with the resulting smaller
fragments more resilient to tidal effects.

Because these asteroids are presumably composed of largely ductile metal, it is
interesting to consider the process of breakup itself. Before catastrophic
mechanical failure when the asteroid reaches the tensile limit, it will first
reach the yield limit. At this point additional stress causes plastic
deformation, leaving the planetesimal shape permanently altered even if tensile
forces are relaxed. Therefore breakup of metallic asteroids will result in
deformation before mechanical failure \citep{slyuta13-1}. \citet{knox70-1}
found yield strengths of 400\,MPa to be typical, with \citet{petrovic01-1}
showing that at room temperature, the ratio of tensile and yield strengths is
typically 1.5--2. This act of deformation may hasten the breakup process, as
the stretching will result in tidal heating, further weakening the metal.
Additional heating/weakening could be provided by flux from the central star,
but this strongly depends on whether the asteroid remains in the vicinity of
the white dwarf long enough for it to thermally respond. While a temperature
dependent reduction in mechanical strength will cause the planetsimal to
disintegrate further away from the white dwarf, the ratio of tensile to yield
strength will increase \citep{petrovic01-1} allowing for a greater degree of
deformation before fragmentation.

Although rather simplified, the size dependence of equation\,\eqref{eq:td}
indicates that the resulting fragments are more resistant to tidal forces, and
so must move closer to the central white dwarf before further breakup can
occur. This logically implies continuous fragmentation down to the surface of
the white dwarf, with some final size for the accreted pieces. Setting the left
hand side of equation\,\eqref{eq:td} to $R_\mathrm{wd}$, this implies km-sized
fragments reaching the white dwarf surface, as also indicated by
\citet{brownetal17-1} in their analysis of granite. In reality this minimum
size will be smaller than 1\,km due to the temperature dependence of the
tensile strength and whether any significant melting/ablation of the fragments
occurs on their final descent. That being said, even if we reduce the tensile
strength by a factor of one hundred, this only reduces the final fragment size
by a factor ten, i.e. fragments on the order of 100\,m arriving at the white
dwarf surface. In other words it is quite possible that the metal-rich material
we see at \sdss{0741}{+}{3146} and \sdss{0823}{+}{0546} did not accrete on to
their respective white dwarfs entirely in the gas phase, but rather impacted
the stellar photosphere as millions of solid fragments.

Because these impacts would occur at orbital speeds of several $1000$\,\kms,
the expected impact energy would be very large -- a 100\,m diameter iron sphere
would impact a canonical 0.6\,\Msun, 0.013\,\Rsun\ white dwarf with a free-fall
energy of $\sim10^{30}$\,erg. Because the scale-heights for cool white dwarfs
with helium dominated atmospheres are on the order of 10\,m, the energy release
will take place on $\mu$s time-scales and thus lead to a short short-lived, but
luminous burst. \citet{brownetal17-1} also considered this situation, with
comparison to Solar impactors, with their analysis suggesting energies direct
collisions of km-sized planetesimal fragments on to white dwarfs could be
observable as transient sources. While we have conservatively assumed smaller
fragments than \citet{brownetal17-1}, the specific energy of 100\,keV per
nucleon is independent of the planetesimal mass\footnote{\citet{brownetal17-1}
quote a specific energy of 10\,MeV per nucleon implying emission up to
gamma-ray energies. However, this is in error and should be 100\,keV per
nucleon, and thus maximum emission energies of hard X-rays (J. Brown, priv.
comm., 2017).}. Therefore the systems discussed in this section motivate future
searches into high energy transients from direct impacts of solid bodies on to
nearby accreting white dwarfs.

\subsection{Mg-rich objects}
\label{mg_rich}

\begin{figure*}
  \centering
  \includegraphics[angle=0,width=\textwidth]{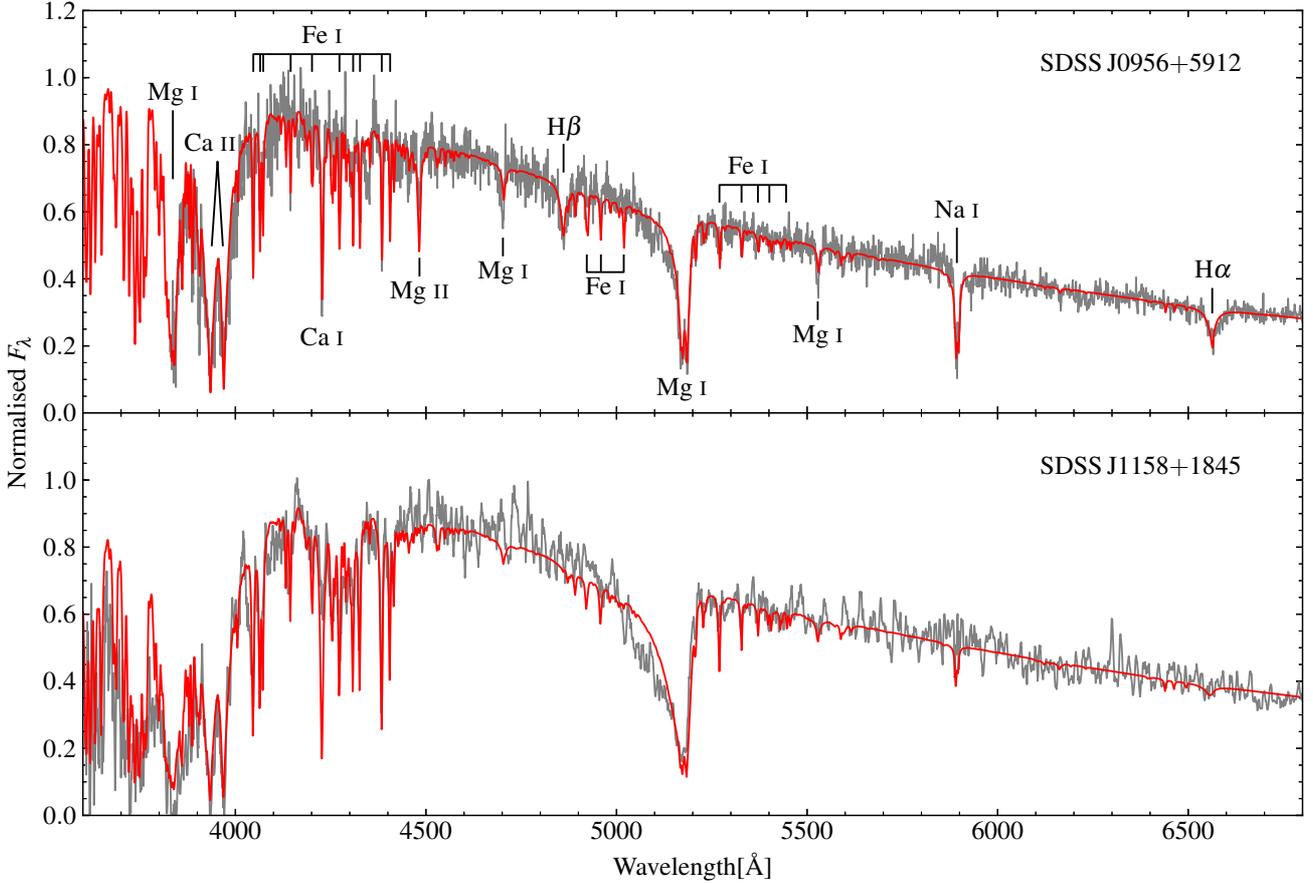}
  \caption{\label{fig:spectra_Mg}
  The two objects classed as Mg-rich show strong absorption from the
  \Ion{Mg}{i}-b triplet located at 5171\,\AA. While \sdss{0956}{+}{5912}
  exhibits Balmer lines, the atmosphere is in fact helium dominated, with
  hydrogen as a trace element.
  }
\end{figure*}

\sdss{0956}{+}{5912} and \sdss{1158}{+}{1845} both stand out in
Fig.~\ref{fig:ternary} as Mg-rich (\sdss{1158}{+}{1845} is the right-most of
the two blue points) and exhibit strong \Ion{Mg}{i} lines in their spectra
(Fig.~\ref{fig:spectra_Mg}), They are are even more Mg-rich than the Earth's
mantle (blue cross), however it is not clear whether these abundances reflect
atypical planetesimal compositions, for example pure magnesium silicate, or the
result of relative diffusion. As outlined in Section~\ref{reldiff}, we find
that for the stars in our sample, the Mg diffusion time-scales are typically
$2.8$ times longer than those of Ca or Fe. Therefore any given point in
Fig.~\ref{fig:ternary} (with the exception of the left-edge, which implies zero
initial Mg) will move towards the bottom right corner over time, as Ca and Fe
diffuse out of the white dwarf CVZ faster than Mg.

Inspection of Fig.~\ref{fig:ternary} shows that a line drawn between the bulk
Earth composition and the bottom right corner passes close to both
\sdss{0956}{+}{5912} and \sdss{1158}{+}{1845}. Therefore, it is perhaps
possible that these white dwarfs may have accreted Earth-like material several
Myr ago which is now severely Mg enhanced. In units of Ca
diffusion-time-scales, $\tau_\mathrm{Ca}$, an initial composition resembling
the bulk Earth will arrive at the present position of \sdss{0956}{+}{5912} in
approximately $2.5\,\tau_\mathrm{Ca}$ ($\simeq 3.2$\,Myr), and about
$3\,\tau_\mathrm{Ca}$ ($\simeq3.7$\,Myr) for \sdss{1158}{+}{1845}. For
\sdss{1158}{+}{1845} this corresponds to an initial Ca abundance close to
$-6.5$\,dex. This slightly surpasses \sdss{1340}{+}{2702} which has the highest
observed Ca abundance in our sample at $-6.98$\,dex. Assuming a $0.6$\,\Msun\
white dwarf, the total accreted mass (scaling from bulk Earth abundances),
would have been roughly that of Ceres. For \sdss{0956}{+}{5912} the situation
is even more extreme. Despite the very high Mg/Ca ratio, the absolute
Ca-abundance is the second highest in our sample. For an accretion episode
occuring $2.5\,\tau_\mathrm{Ca}$ ago, the original Ca abundance would have been
about $-6.1$\,dex, corresponding to total accreted metals of about
$3\times10^{24}$\,g or 3 Ceres masses (again assuming a 0.6\,\Msun white
dwarf).

As very few objects within the Solar system have masses in this range and
above, it seems unlikely that we observe two systems having accreted such
extremely large planetesimals. An alternative hypothesis is that increased
dynamical activity at these systems several Myr ago led to the accretion of a
large number of lower mass planetesimals totalling a Ceres mass or more. For
instance tidal interactions of passing stars could provide such short lived
dynamic instabilities \citep{bonsor+veras15-1,hamersetal16-1,verasetal17-1}.
Several Myr later after this intense accretion episode has ceased, these white
dwarfs show Mg-rich material due to relative diffusion.

Finally we consider the possibility that the parent bodies accreted by
\sdss{0956}{+}{5912} and \sdss{1158}{+}{1845} were intrinsically Mg-rich, far
from our assumption that planetesimals follow the core/mantle/crust
compositions of the Earth. For instance, the observations can be considered
consistent with parent body masses of $\sim 10^{23}$\,g accreted much more
recently, but with substantially higher Mg/Ca and Mg/Fe ratios than even the
Earth's mantle, e.g. predominantly enstatite, forsterite, or a mixture of the
two.

\section{Evolution of remnant planetary systems}
\label{ages}

\Teff\ and atmospheric abundances are determined directly from spectral
fitting, and so it is common practice to plot the abundance of one of the
metals (usually Ca, as this is most easily detected) against \Teff. While this
is a useful way to illustrate variations in abundances across the full range of
known white dwarfs, the non-linear relationship between \Teff\ and age does not
provide the best handle on the evolution of the oldest systems. For instance
Fig.~8 of \citet{koesteretal14-1} shows that the highest observed accretion
rates ($10^9$\,g\,s$^{-1}$) of rocky debris on to DA white dwarfs remain
constant over a large range in \Teff, however most of the sample discussed in
that paper spans only the first Gyr of white dwarf cooling. Several dynamical
studies suggest that a decrease in scattering (and subsequent accretion) events
ought to occur \citep{debesetal12-1, mustilletal14-1, verasetal13-1,
verasetal16-1}. In this section we provide evidence for a decline in maximum
observed accretion rate, but over time-scales of many Gyr, which we are able to
probe for the first time with our sample of cool DZ white dwarfs.

Since white dwarfs cool predictably after departing the AGB, we were able to
estimate cooling ages for our DZ sample (Paper~I). White dwarf ages depend not
only on the white dwarf \Teff, but also the white dwarf mass. As it is not
possible to determine these spectroscopically for DZs, we used the SDSS white
dwarf mass distribution \citep{kepleretal15-1} as a prior. Propagating the
measured \Teff\ values and mass prior through the Montreal DB cooling tracks
\citep{bergeronetal01-1,holberg+bergeron06-1,kowalski+saumon06-1,bergeronetal11-1},
we thus were able to determine cooling ages and their associated uncertainties.
A more detailed description of these cooling age calculations are given in
Paper~I, with the results tabulated in Table~8 of that work.

The oldest system in our sample is \sdss{1636}{+}{1619} at
$7.7^{+0.3}_{-0.9}$\,Gyr, which is unsurprising considering it also has the
reddest $g-r$ colour at $1.10\pm0.03$ (See paper I). Because the typical white
dwarf progenitor is a $\simeq2$\,\Msun\ A-type star \citep{catalanetal08-2}
with a main-seqence life time of $\sim1$\,Gyr, the total system age is
$\simeq9$\,Gyr. Therefore, our analysis indicates this object is nearly as old
as the Galactic disc \citep{oswaltetal96-1,delpelosoetal05-1,haywoodetal13-1},
yet still shows the signs of a planetary system, long after departing the
main-sequence.

\begin{figure}
  \centering
  \includegraphics[angle=0,width=\columnwidth]{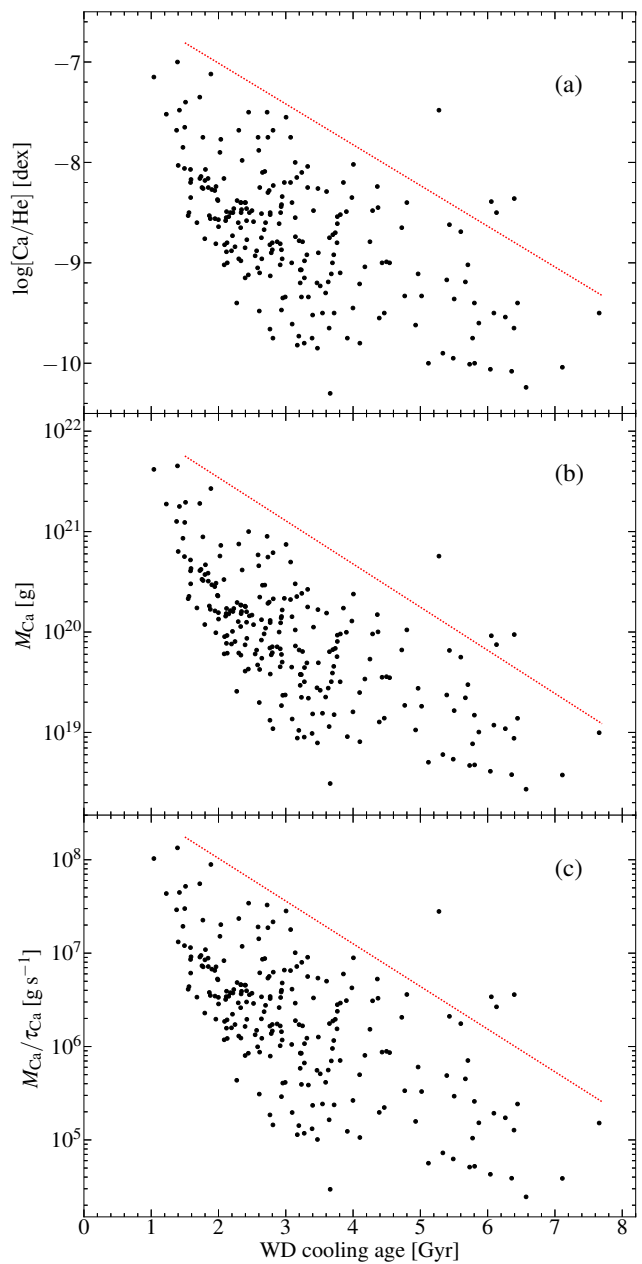}
  \caption{\label{fig:ages}
    (a) Atmospheric Ca abundances against white dwarf age. The dotted line
    indicates our inferred upper bound to the distribution with the exception
    of a few outliers (which are discussed in section~\ref{outliers}).
    (b) The Ca abundances have been rescaled by the CVZ masses, yielding the
    mass of Ca in the white dwarf CVZs. The slope of the dotted line is
    adjusted accordingly.
    (c) Ca mass is divided by the time-scale for Ca to diffuse out of the
    bottom of the CVZ. This can be interpreted as a diffusion flux or mean
    accretion rate. Again, the slope of the dotted line is updated.
    }
\end{figure}

In Fig.~\ref{fig:ages}a we show \logX{Ca} versus the estimated cooling ages.
The distribution is approximately triangular in shape, however only one of the
edges has a physical significance. The left and lower edges (young and less
polluted systems) merely result from our white dwarf identification method,
which by design is insensitive to hot/young systems or white dwarfs with very
low metal-abundances (Paper~I). In contrast, the upper edge is an apparently
real boundary of the DZ white dwarf distribution, representing a decrease in
the maximum-encountered Ca abundance of $\approx 2.5$\,dex across the full
age-range of our sample.

Selection-bias can be easily ruled out as hypothetical objects within the
upper-right corner (cool and extremely metal-rich), would have highly
distinctive spectra, and would look unlike any main-sequence star or quasar.
The corresponding model spectra were calculated for the grid used in Paper~I
for the search of cool DZs. Therefore, we are confident that we would have
identified any such system. The absence of objects with $\logX{Ca} \simeq
-7$\,dex at old ages indicates they must be extremely rare\footnote{A clear
exception is \sdss{0916}{+}{2540} with $\logX{Ca} = -7.5$\,dex and cooling age
of $5.3_{-1.2}^{+0.6}$\,Gyr, which we discuss in Section~\ref{outliers}.}.

A physical interpretation for the decrease of $\logXY{Ca}{He}$ must account for
the $2.5$\,dex change we see in Fig.~\ref{fig:ages}a. We show here that neither
variations in the size of the white dwarf CVZ, nor elemental diffusion
time-scales are significant enough to explain the magnitude of this decrease,
and thus we are unable to explain the abundance decrease as the result of
evolving white dwarf properties.

For each white dwarf in our sample, we performed envelope calculations to
determine the masses of their outer CVZs as well as the diffusion time-scales
for each element (Paper I, Table~7). These properties change with white dwarf
cooling, and so we investigated whether these could account for the trend seen
in Fig.~\ref{fig:ages}a. We firstly scaled Ca abundances by the masses of the
white dwarfs CVZs, which determines the mass of Ca mixed throughout the white
dwarf envelopes. This had little effect on the trend, which remained at a
$\simeq2.5$\,dex decrease across the age range of our sample
(Fig.~\ref{fig:ages}b). We then rescaled these masses by each white dwarf's Ca
diffusion time-scale, which determines the mass fluxes through the base of
their CVZs, or in other words, the average accretion rates of Ca on to the
stars. Rather than causing the trend to subside, we instead found it to steepen
to $\simeq 3$\,dex (Fig.~\ref{fig:ages}c)

The downwards trend is indicated by the dashed lines in each panel of
Fig.~\ref{fig:ages} and corresponds to an exponential decrease in the accretion
rate upper bound with white dwarf age (for now ignoring four outliers above the
lines, discussed separately below). The slope of the line in
Fig.~\ref{fig:ages}c corresponds to an e-folding time-scale of 0.95\,Gyr. We
found that this could not be varied much more than 0.1\,Gyr before appearing
incompatible with the data, and thus we argue that $0.95\pm0.10$\,Gyr is the
time-scale on which the accretion rate upper limit decays for our DZ sample.
Since this decrease of white dwarf pollution with age does not appear to arise
from either selection bias nor a change white dwarf properties, it likely
relates to the properties of the planetary systems at these white dwarfs.

We find the most reasonable explanation is that the number of planetesimals
remaining in old remnant planetary systems, available to be scattered towards
the white dwarf, decreases with time. Since the occurrence rate that white
dwarfs accrete planetesimals will be proportional to the number available to be
scattered inwards, then an exponential decrease in the largest objects is to be
expected. Dynamical simulations have previously suggested that the occurrence
rate of white dwarf pollution should be expected to decrease on Gyr time-scales
\citep{debesetal12-1,mustilletal14-1,verasetal13-1,verasetal16-1} We therefore
suggest that our observations may show the first evidence of this process
occurring.

Our results in Paper~I provide one caveat to this interpretation. In Fig.~11 of
that work, we compared our DZs with the DZ sample of \citet{dufouretal07-2} and
the DBZ sample of \citet{koester+kepler15-1}. There we noted an abrupt 2\,dex
increase in \logX{Ca} occurring at about 10\,000\,K (corresponding to a cooling
age of 0.7\,Gyr for a 0.6\,\Msun\ white dwarf), and speculated that this may
indicate an incomplete understanding of white dwarf CVZ formation. It is
therefore prudent to remain cautious of the 3\,dex decrease we see here in
Fig.~\ref{fig:ages}. On the other hand, our envelope calculations currently
suggest that the combined effect of variations in CVZ sizes and diffusion
time-scales across our sample act to amplify the decline between
Figs.~\ref{fig:ages}a and \ref{fig:ages}c, although only by about 0.5\,dex.
Therefore a change in our understanding of white dwarf CVZs would need to imply
a three order of magnitude change in the opposite direction to remove the trend
seen in Fig.~\ref{fig:ages}c. 

A hypothesis for the abrupt rise in $\logX{Ca}$ (Fig.~11 of Paper I), that we
did not consider in that work, is that a dynamical instability occurs after
$\simeq0.7$\,Gyr of white dwarf cooling ($\Teff\simeq10\,000$\,K),
spontaneously increasing the occurrence rate at which planetesimals are
accreted \citep[e.g. see Fig.~3 of ][]{mustilletal14-1}. For white dwarfs with
hydrogen atmospheres with their short sinking time-scales, such an instability
would instead manifest itself as an increase in the DAZ/DA ratio at around the
same age which, subject to selection biases, is potentially observed in Fig.~8
(middle panel) of \citet{koesteretal14-1}, Thus the changes in maximum
abundance seen in Fig.~11 of Paper~I and Fig.~\ref{fig:ages} here could both be
related to the evolution of their planetary systems.

\subsection{Metal rich outliers}
\label{outliers}

\begin{figure*}
  \centering
  \includegraphics[angle=0,width=\textwidth]{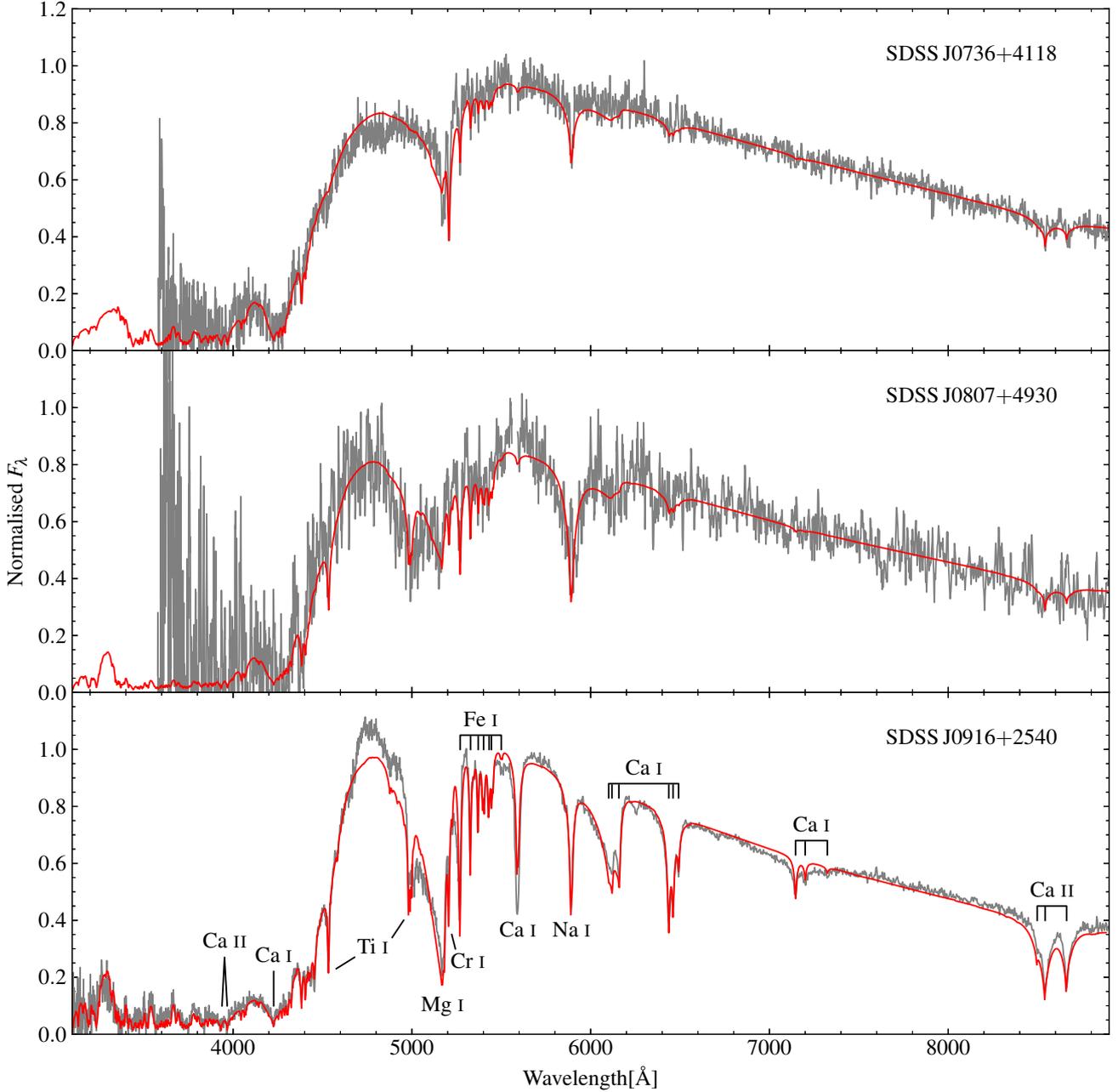}
  \caption{\label{fig:spectra_0916}
  Spectra for three of the four outliers in Fig.~\ref{fig:ages} are shown with
  their best fitting atmospheric models. The remaining system,
  \sdss{0744}{+}{4649} is already shown in Fig.~\ref{fig:spectra_Ca}. The
  unique spectrum of \sdss{0916}{+}{2540} exhibits many deep, broad absorption
  features with some of the shallower lines observed only in this DZ.
  }
\end{figure*}

We note that four systems in our sample (\sdss{0736}{+}{4118},
\sdss{0744}{+}{4649}, \sdss{0807}{+}{4930}, and \sdss{0916}{+}{2540}) are
located above the upper envelope for all three parametrizations in
Fig.\,\ref{fig:ages}. Naturally these outliers are worthy of discussion in
regards to their unusually high Ca-abundances/accretion rates for their ages.
The spectra of \sdss{0736}{+}{4118}, \sdss{0807}{+}{4930}, and
\sdss{0916}{+}{2540} are displayed in Fig.~\ref{fig:spectra_0916} -- the
spectrum of \sdss{0744}{+}{4649} can be found in Fig.~\,\ref{fig:spectra_Ca}. 

\smallskip
\noindent\textbf{\sdss{0744}{+}{4649}}\\
This system has already been discussed in detail in Section~\ref{ca_rich}, due
to the Ca-rich nature of the accreted material which, combined with the
moderate total-metal-abundance for this star, leads to a particularly high
location in Fig.~\ref{fig:ages}. In terms of \logX{Mg} or \logX{Fe}, this star
is located within the distribution of our sample (although only just).

\smallskip
\noindent\textbf{\sdss{0916}{+}{2540}}\\
This extremely metal-rich DZ shows a spectrum quite unlike any other white
dwarf, with extreme photospheric absorption by a large number of elements
across the entire optical range. This system was first analysed by
\citet{koesteretal11-1}\footnote{ \sdss{0916}{+}{2540} was also independently
identified by members of the Galaxy Zoo community from its SDSS spectrum, who
also correctly classified it as an unusual white dwarf
(\url{http://www.galaxyzooforum.org/index.php?topic=276688.15}).} in their
sample of 26 DZ stars. As our DZ sample is nine times larger than that of
\citet{koesteretal11-1}, one might expect to find several more similar objects,
however \sdss{0916}{+}{2540} remains unique among our 230 objects. No doubt
such an usual spectrum could be recognised even in low quality data (but with
reduced scope for spectral analysis), and so we are lucky that this
intrinsically faint star is so nearby ($d=43.4\pm5.4$\,pc) that its spectrum
can be studied in exquisite detail.

Within Fig.~\ref{fig:ages}, \sdss{0916}{+}{2540} is an order of magnitude more
abundant in Ca, compared to other white dwarfs of similar temperature/age,
suggesting some rare phenomenon results in its extremely metal-rich
photosphere. The answer, we believe, lies approximately 40 arcseconds to the
South-East in the form of a K-star common-proper-motion companion (J. Farihi,
priv. comm., 2013). At the estimated distance of $43.4\pm5.4$\,pc (Paper~I),
the projected separation between the two stars is $1900\pm200$\,au.

Several recent theoretical studies suggest that wide binary companions can
cause secular instabilities in white dwarf planetary systems, even at large
ages, resulting in an increased influx of planetesimals.
\citet{bonsor+veras15-1} considered the effect of Galactic tides on wide binary
systems, and found that secular variations in the orbital elements, can lead to
a close approach several Gyr after the primary has entered the white dwarf
cooling track, thus turning a previously stable planetary system into a
dynamically active one.

In contrast, \citet{petrovich+munoz17-1} considered systems where a belt of
exoplanetesimals is initially located between an inner planetary system and a,
potentially stellar, inclined outer companion. During the main-sequence, they
argued that the inner planetary system dynamically shields planetesimals
against perturbations from the outer companion. However if the inner system is
engulfed during stellar evolution to the white dwarf stage, then the
planetesimal belt that may have been dynamically stable throughout the
main-sequence can now be affected by the outer perturber via the Kozai-Lidov
mechanism. Under this mechanism, an orbiting companion inclined with the plane
of the planetary system causes planetesimals on previously circular orbits to
exchange eccentricity with their own inclination. This can result in
planetesimals on highly eccentric orbits, with pericentres within the white
dwarf tidal-disruption radius. According to \citet{petrovich+munoz17-1}, this
mechanism also has the advantage that it is always active, and so could explain
the extreme abundances at \sdss{0916}{+}{2540}, despite the present binary
separation of several thousand au.

\smallskip
\noindent\textbf{\sdss{0807}{+}{4930}}\\
We note that qualitatively, the spectrum of this white dwarf resembles that of
\sdss{0916}{+}{2540}, but far less extreme (an ``0916-lite''), motivating us to
check this system for binarity. Due to the faintness of this white dwarf
($r=20.5$), no published proper-motion is available, however we were able to
calculate a moderate proper-motion of $\vec{\mu} = (\mu_\alpha\cos\delta,
\mu_\delta) = (-113.9\pm3.3, -54.0\pm3.1)$\,\masy using two imaging epochs from
SDSS (2000.3154 and 2003.8123), and one from Pan-STARRS (2013.5689). Searching
nearby stars for binary membership revealed an obvious companion 27 arcseconds
to the North-West. Based on its colours, the companion is a mid-to-late M-type
star ($r=20.1$) with proper-motion $\vec{\mu} = (-112.6\pm5.6,
-51.1\pm5.6)$\,\masy. At an estimated distance of $156\pm20$\,pc (Paper~I),
this implies a projected separation of $4200\pm500$\,au.

While 70\,percent of the DZs in our sample have proper motion measurements, we
find no evidence for wide companions to any of the white dwarfs further to
\sdss{0807}{+}{4930} and \sdss{0916}{+}{2540}. Therefore these two systems
provide a strong case that binarity is correlated with higher than average
accretion rates. We also note that WD\,1425+540, recently analysed by
\citet{xuetal17-1}, is also a member of wide binary, where the companion is
speculated to have provided the perturbations leading to the accretion of a
Kuiper-belt-like object by the white dwarf. In the context of systems like
WD\,1425+540, \citet{stephanetal17-1} also found the Kozai-Lidov mechanism as a
feasible way to scatter long-period objects (i.e. Kuiper-belt analogue objects)
towards white dwarfs. While we find no evidence of volatile elements (C, N, S)
in the atmospheres of either \sdss{0807}{+}{4930} or \sdss{0916}{+}{2540},
\citet{stephanetal17-1} found that wide-companions raise the likelihood of
accretion from icy-bodies, particularly for low-mass companions as with the two
wide binaries we observe here.

\smallskip
\noindent\textbf{\sdss{0736}{+}{4118}}\\
Of the four outliers, \sdss{0736}{+}{4118} is the only system with no obvious
property naturally explaining its high Ca abundance in Fig.~\ref{fig:ages}.
However of the four objects, it is also the least extreme ($t_\mathrm{cool} =
6.1$\,Gyr, $\logX{Ca}=-8.50$\,dex). Therefore it may simply be the case that
\sdss{0736}{+}{4118} has accreted its atmospheric metals much more recently
than other systems in our sample or even that accretion is still ongoing.

\section{Conclusions}
\label{conclusions}

The cool white dwarf sample that we first presented in
\citet{hollandsetal17-1}, contains a wealth of information on the compositions
and evolution of planetary objects around white dwarfs.

The white dwarf spectra show lines of Ca, Mg, and Fe, and in some cases also
Na, Cr, Ti or Ni. Relative abundances for each of these elements demonstrate
that the compositions of exoplanetary debris is diverse. In particular we
identify clear signs of differentiation in several systems, with abundance
patterns closely matching crust- and core- like material. Using a simple
technique to estimate mass-fractions of crust/mantle/core material, we show
that our most extreme systems, as well as 22 exoplanetesimals compiled from the
literature range from crust-like to core-like with mixed crust+mantle+core
systems in between, but there is, as expected, a dearth of systems consistent
with crust+core but lacking a mantle component. From this analysis we also
conclude that two of our systems, \sdss{0741}{+}{3146} and
\sdss{0823}{+}{0546}, have accreted the most core-like exoplanetesimals
discovered to date.

This sample of cool white dwarfs also spans a wide range in cooling ages from 1
to almost 8\,Gyr. We show that the diminishing amount of metal pollution with
increasing age may provide evidence for the slow decay in the sizes of remnant
planetary systems as the largest planetesimals are scattered away. This occurs
on an e-folding time-scale of about $0.95\pm0.10$\,Gyr, an effect that
previously escaped notice due to selection bias towards younger systems that do
not sufficiently sample changes on this time-scale.

Finally we find that the only two confirmed binary members in our sample,
exhibit enhanced metal abundances compared to other objects of similar
cooling-ages. This provides convincing evidence that white dwarfs in binary
systems experience higher than average accretion rates of exoplanetesimals, as
a result of dynamical instabilities that have been previously argued for
theoretically.

\section*{Acknowledgements}

MH acknowledges useful discussions with Amy Bonsor on the lifetimes of remnant
planetary systems, Oliver Shorttle on iron to calcium ratios, and Dimitri Veras
for many insightful conservations on the breakup of asteroids with high tensile
strength. MH also acknowledges funding from the University of Warwick
Chancellor's fellowship. Ben Zuckerman is acknowledged for providing valuable
comments to the benefit of the manuscript. The anonymous referee is
acknowledged for useful feedback which greatly improved the introduction to
this work. The research leading to these results has received funding from the
European Research Council under the European Union's Seventh Framework
Programme (FP/2007-2013) / ERC Grant Agreement n. 320964 (WDTracer). Funding
for the Sloan Digital Sky Survey IV has been provided by the Alfred P. Sloan
Foundation, the U.S. Department of Energy Office of Science, and the
Participating Institutions. SDSS- IV acknowledges support and resources from
the Center for High-Performance Computing at the University of Utah. The SDSS
web site is www.sdss.org. This work makes use of observations made with the
William Herschel Telescope operated on the island of La Palma by the Isaac
Newton Group in the Spanish Observatorio del Roque de los Muchachos of the
Instituto de Astrofisica de Canarias.

%%%%%%%%%%%%%%%%%%%%%%%%%%%%%%%%%%%%%%%%%%%%%%%%%%



\bibliographystyle{mnras}
\bibliography{aamnem99,aabib}

%\begin{thebibliography}{99}
%\end{thebibliography}

%%%%%%%%%%%%%%%%%%%%%%%%%%%%%%%%%%%%%%%%%%%%%%%%%%

%%%%%%%%%%%%%%%%% APPENDICES %%%%%%%%%%%%%%%%%%%%%

%\appendix
%\onecolumn

%%%%%%%%%%%%%%%%%%%%%%%%%%%%%%%%%%%%%%%%%%%%%%%%%%

% Don't change these lines
\bsp	% typesetting comment
\label{lastpage}
\end{document}